\documentclass{aastex62}

\usepackage{booktabs}

\def\kms  {km~s$^{-1}$}

\received{October 26, 2019}
\revised{April 1, 2020}
\accepted{April 18, 2020}
\submitjournal{ApJS}

\shorttitle{44 GHz methanol masers}
\shortauthors{Yang et al.}

\begin{document}

\title{44 GHz methanol masers: Observations toward 95 GHz methanol masers}

\correspondingauthor{Wenjin Yang}
\email{wjyang@pmo.ac.cn, xuye@pmo.ac.cn}

\author[0000-0002-3599-6608]{Wenjin Yang}
\affiliation{Purple Mountain Observatory, Chinese Academy of Science, Nanjing 210008, China}
\affiliation{University of Science and Technology of China, 96 Jinzhai Road, Hefei 230026, China}

\author{Ye Xu}
\affiliation{Purple Mountain Observatory, Chinese Academy of Science, Nanjing 210008, China}

\author{Yoon Kyung Choi}
\affiliation{ Max-Planck-Institut f\"{u}r Radioastronomie, Auf dem H\"{u}gel 69, D-53121 Bonn, Germany}

\author[0000-0002-1363-5457]{Simon P. Ellingsen}
\affiliation{School of Natural Sciences, Private Bag 37, University of Tasmania, Hobart, Tasmania, Australia}

\author{Andrej M. Sobolev}
\affiliation{ Ural Federal University, Lenin Ave. 51, Ekaterinburg 620000, Russia}

\author{Xi Chen}
\affiliation{Center for Astrophysics, GuangZhou University, Guangzhou 510006, China}

\author{Jingjing Li}
\affiliation{Purple Mountain Observatory, Chinese Academy of Science, Nanjing 210008, China}

\author{Dengrong Lu}
\affiliation{Purple Mountain Observatory, Chinese Academy of Science, Nanjing 210008, China}



\begin{abstract}
We report a simultaneous 44 and 95 GHz class I methanol maser survey toward 144 sources from the 95 GHz class I methanol maser catalog. The observations were made with the three telescopes of the Korean very long baseline interferometry network operating in single-dish mode. The detection rates are 89\% at 44 GHz and 77\% at 95 GHz. There are 106 new discoveries at 44 GHz. 
Comparing the previous 95 GHz detections with new observations of the same transitions made using the Purple Mountain Observatory 13.7 m radio telescope shows no clear evidence of variability on a timescale of six years. 
Emission from the 44 and 95 GHz transitions shows strong correlations in peak velocity, peak flux density, and integrated flux density, indicating that they are likely cospatial. 
We found that the peak flux density ratio $S_{\rm pk,95}$/$S_{\rm pk,44}$ decreases as the 44 GHz peak flux density increases.
We found that some class I methanol masers in our sample could be associated with infrared dark clouds, while others are associated with H {\sc ii} regions, indicating that some sources occur at an early stage of high-mass star formation, while others are located toward more evolved sources.
\end{abstract}

\keywords{masers --- ISM: molecules ---  radio lines: ISM --- star: formation}


\section{Introduction} \label{sec:intro}

Methanol masers are commonly found in star formation regions (SFRs).  The methanol molecule has numerous transitions within the centimeter- and millimeter-wavelength ranges and hence provides a powerful tool for the study of star-forming regions.
According to the empirical scheme of \cite{1991ASPC...16..119M}, methanol masers are divided into two categories. 
Class~II methanol masers, such as the 6.7 GHz (5$_{1}$--6$_{0}$A$^{+}$) and 12.2 GHz (2$_{0}$--3$_{-1}$E) transitions \citep{1987Natur.326...49B,1991ApJ...380L..75M}, are found in close proximity to ultracompact H {\scriptsize II} regions, infrared sources, and OH masers \citep{1998MNRAS.301..640W}, and are known to be pumped by radiative processes \citep{1992MNRAS.259..203C,2005MNRAS.360..533C}.
In contrast, class I methanol masers are thought to be tracers of shocked regions related to outflows and expanding H{\sc ii} regions \citep[e.g.,][]{2013ApJ...763....2G,2014MNRAS.439.2584V} and are produced by collisional pumping. 
On the basis of the latest class I methanol maser models \citep{2016A&A...592A..31L,2018IAUS..336...57S}, such masers can be divided into several families: the (J+1)$_{0}$--J$_{1}$-A type, the (J+1)$_{-1}$--J$_{0}$-E type, the J$_2$--J$_1$-E series at 25 GHz, and the J$_{-2}$--(J$-$1)$_{-1}$-E series at 9 GHz.
Within the Milky Way, methanol maser emission from the 44 GHz 7$_{0}$--6$_{1}$A$^{+}$ \citep[e.g.,][]{1985ApJ...288L..11M,1992AZh....69.1002K,1994MNRAS.268..464S} is the most widespread and strongest transition, 
with the 36 GHz 4$_{-1}$--3$_{0}$ E \citep[e.g.,][]{1989ApJ...339..949H,1996AA...314..615L}, 84 GHz 5$_{-1}$--4$_{0}$ E \citep[e.g.,][]{2001ARep...45...26K,2019MNRAS.484.5072B}, and 95 GHz 8$_{0}$--7$_{1}$A$^{+}$ \citep[e.g.,][]{2011ApJS..196....9C,2013ApJS..206....9C} transitions also being relatively common.

The first detection of 44 GHz methanol masers was reported by \cite{1985ApJ...288L..11M}.
Subsequent single-dish surveys searching for 44 GHz methanol masers have targeted galactic SFRs \citep{1990ApJ...354..556H}, water masers \citep{1990AA...240..116B}, cold Infrared Astronomical Satellite (IRAS) sources \citep{1992AZh....69.1002K}, H {\sc ii} regions \citep{1994MNRAS.268..464S}, 6.7 GHz methanol masers \citep{2010AA...517A..56F}, intermediate-mass young stellar objects \citep[YSOs;][]{2011ApJS..196...21B}, and 2.122 $\mu$m emission \citep{2012AJ....144..151L}. These searches have laid the crucial groundwork for establishing the nature of the 44 GHz methanol maser transition.
Recently, a series of interferometric observations targeting outflows \citep{2016ApJS..222...18G} and high-mass protostellar objects \citep[HMPOs;][]{2017ApJS..233....4R}, have revealed the relationship between the class I methanol masers and other star-forming phenomena at high angular resolution, which has allowed more detailed investigations of the evolutionary stage and the maser excitation conditions.
In addition, variability of class I masers has only been the subject of limited studies, and to date, no detectable variability has been reported \citep[e.g.,][]{2012ApJ...759..136C}. 

The 8$_{0}$--7$_{1}$A$^{+}$ transition of methanol at 95 GHz is a member of the same transition family as the 44 GHz maser transition.
Systematic searches for 95 GHz methanol masers toward Spitzer Galactic Legacy Infrared Mid-Plane Survey Extraordinaire (GLIMPSE) extended green objects \citep[EGOs;][]{2011ApJS..196....9C,2013ApJS..206....9C}, molecular outflow sources \citep{2013ApJ...763....2G}, cross-matching between the GLIMPSE point sources and the Bolocam Galactic Plane Survey (BGPS) sources \citep{2012ApJS..200....5C}, and BGPS sources \citep{2017ApJS..231...20Y} have significantly increased the number of known 95 GHz methanol masers. \cite{2017ApJS..231...20Y} compiled a catalog summarizing the past two decades of searches for 95 GHz methanol masers.  This catalog contains 481 maser sources and a further 37 candidates, and it currently represents the largest and most complete catalog of 95 GHz methanol masers.

In this paper, we report a search for 44 GHz methanol masers targeting known 95 GHz masers.  This search was made using the three 21 m antennas from the Korean very long baseline interferometry network (KVN) in single-dish telescope mode. In Section \ref{sec:obs} we describe the source selection and observations, then we give the results of the detections, compare them with the previous observations in Section \ref{sec:results}, and follow this by discussing the flux density ratio between the two maser transitions and their occurrence at different stages of the star formation process (Section \ref{sec:discuss}) and a summary in Section \ref{sec:sum}.

\section{Source Selection and Observations} \label{sec:obs}

\subsection{Source selection} \label{sec:source}
The 95 GHz methanol maser sources in which we searched for emission from the 44 GHz methanol transition were selected using the following criteria: 
(1) The sources are listed in the catalog of 95 GHz methanol masers and maser candidates \citep{2017ApJS..231...20Y}.
The \citet{2017ApJS..231...20Y} catalog of 95 GHz methanol masers contains 481 sources and 37 maser candidates, identified in searches toward BGPS sources \citep[version 1.0.1;][]{2010ApJS..188..123R} and other searches undertaken over the past two decades \citep{1994AAS..103..129K,1995AZh....72...22V,2000MNRAS.317..315V,2005MNRAS.359.1498E,2006ARep...50..289K,2010AA...517A..56F,2011ApJS..196....9C,2012ApJS..200....5C,2013ApJS..206....9C,2013ApJ...763....2G}. 
Sources in which only broad spectral components \citep[$>$ 2.5 \kms;][]{2017ApJS..231...20Y} are observed at 95 GHz were considered maser candidates.
(2) Sources with declination greater than $-$25$^\circ$ (which are observable with the KVN 21m antennas). 
(3) Sources with previously reported 44 GHz maser detections from single-dish observations were excluded \citep{1990ApJ...354..556H,1990AA...240..116B,1992AZh....69.1002K,1994MNRAS.268..464S,2006ARep...50..289K,2008AJ....135.1718P,2010AA...517A..56F,2011ApJS..196...21B,2012AJ....144..151L,2015ApJS..221....6K}. 
(4) Sources with previous KVN observations, separated by less than 30$\arcsec$ from a \citet{2017ApJS..231...20Y} 95 GHz source from targeting of class~I masers \citep[private communication;][]{1994ApJS...91..659K,1998AA...336..339M,2005AA...432..737P,2006ApJ...641..389R,2007ApJ...656..255P,2008AJ....136.2391C,2010MNRAS.404.1029C,2010MNRAS.409..913G} are also excluded. 
Based on these criteria, a total of 144 95 GHz methanol maser sources were identified (see Table \ref{tab:144_obs}). 
Of these, 123 ($\sim$ 85\%) are sources selected from \cite{2012ApJS..200....5C} and \cite{2017ApJS..231...20Y} that target the position of peak emission of associated BGPS sources \citep[version 1.0.1;][]{2010ApJS..188..123R} whose beam-averaged H$_2$ column density is greater than 10$^{22.1}$ cm$^{-2}$, the others are targeting star-forming regions or outflow/EGO sources with a 95 GHz methanol maser detection. Ten of the 144 target sources are 95 GHz methanol maser candidates, and these are identified with a double dagger in Table~\ref{tab:144_obs}.

\subsection{KVN observations} \label{sec:kvnobs}
The observations of the 44.06943 GHz (7$_{0}$--6$_{1}$A$^{+}$) and 95.16946 GHz (8$_{0}$--7$_{1}$A$^{+}$) class I methanol maser transitions took place in the period 2016 October to December using the KVN 21m radio antennas in single-dish telescope mode.  There are three KVN stations: KVN Yonsei telescope (Seoul), KVN Ulsan telescope (Ulsan), and KVN Tamna telescope (Jeju island). 
Each of the KVN antennas are equipped with a multifrequency receiving systems able to simultaneously operate in the 43 and 86 GHz bands, which enables observing 44 and 95 GHz methanol transitions at the same time. The beam size and antenna efficiency of each antenna are listed in Table \ref{tab:tele_para}.
Telescope pointing observations were performed at least once an hour, resulting in a pointing accuracy better than $\sim$6$\arcsec$.
Digital spectrometers with 4096 spectral channels were used as back-ends, each with a bandwidth of 64 MHz, which provides velocity resolutions of 0.11 and 0.05 \kms at 44 GHz and 95 GHz, respectively. 

A low-order polynomial baseline was fit to the line-free channels in the spectra.   
Hanning smoothing was applied once (twice) to improve the signal-to-noise ratio for the 44 GHz (95 GHz) methanol line spectra.  This results in a velocity-channel width of 0.21 and 0.20 \kms\/ for the 44 and 95 GHz transitions, respectively.
The system temperature varied between 150 K and 300 K depending on the weather conditions and the elevation of the telescope. 
The on-source integration times were between 5 and 15 minutes, which achieved a typical 1$\sigma$ noise level of between 0.4 and 1 Jy for both of the methanol transitions at the $\sim$0.2 \kms\/ resolution obtained after smoothing.
The data were reduced and analyzed using the GILDAS/CLASS package \citep{2005sf2a.conf..721P}. 
To characterize the spectra, we undertook Gaussian fitting of each peak in the spectrum for each source.

Twenty-four sources which were observed in bad weather and were not reobserved.  These sources are indicated by a dagger in Table \ref{tab:144_obs}. 
These sources may have unreliable flux densities for any detected emission, and they have been excluded from our analysis of intensity ratios in Section \ref{sec:95_compare}, \ref{sec:peak_ratio},   \ref{sec:iras}, \ref{sec:sec_irdc}, and \ref{sec_hii}.

\subsection{PMO observations}

The current KVN search did not detect 95 GHz methanol emission toward a small number of targets for which detections of this transition have been reported previously (see Section \ref{sec:kvn_results} for details).
The majority of 95 GHz methanol masers/candidates as KVN targets were detected using the Purple Mountain Observatory (PMO) 13.7m radio telescope with a typical on-source integration time of 10--20 minutes \citep{2017ApJS..231...20Y}, and have a peak flux density greater than 4~Jy.
To eliminate potential effects due to different beam sizes and effectively determine whether these 95 GHz methanol masers show intrinsic intensity variability, we reobserved 13 sources using the PMO 13.7m radio telescope. 
All the 13 sources (see Table \ref{tab:f_comparison}) were observed using the PMO 13.7m in previous studies with a 95~GHz peak flux density stronger than 4 Jy.
Nine of them were not detected in the current KVN 21m observations, the other four sources were randomly selected among the KVN detections.

Observations toward thirteen 95 GHz class I methanol maser sources were made with the PMO 13.7m radio telescope in Delingha, China on 2018 December 4 and 6.
A 3$\times$3 multibeam sideband-separating superconducting spectroScopic array receiver system was used to observe the 95 GHz methanol transition. 
This receiver operates over a frequency range of 85--115 GHz, and the beam in the middle of the first row of the receiver was pointed at the target position in most cases. 
The spectra were recorded using a fast Fourier transform spectrometer with 16,384 spectral channels across a bandwidth of 1 GHz, yielding a frequency resolution of 61 kHz and an effective velocity resolution of 0.19 km s$^{-1}$ for the 95 GHz methanol transition. 
The system temperature for the 95 GHz methanol maser observations was in the range 135--200 K, depending on the weather conditions and telescope elevation. Most sources were observed in a position-switching mode with an off-position offset by 15$\arcmin$ in R.A.  
For some sources, a different reference position was chosen to ensure that the reference spectrum was free from emission. 
The pointing rms was better than 5$\arcsec$. 
The standard chopper-wheel calibration technique \citep{1976ApJS...30..247U} was applied to measure the antenna temperature, $T^{*}_{\rm A}$, corrected for atmospheric absorption. 
The beam size of the telescope is approximately 55$\arcsec$ at 95 GHz, with a main-beam efficiency $\eta_{\rm mb}$ of 62\%. 
The antenna efficiency is 48\%, corresponding to a factor of 39.0 Jy K$^{-1}$ to convert antenna temperature into flux density. 
An on-source integration time of 15 minutes for each source achieved rms noise levels of approximately 1.4 Jy for the 95 GHz methanol transition. 

The data were reduced and analyzed using the GILDAS/CLASS package \citep{2005sf2a.conf..721P}. The data reduction procedure was similar to that of the KVN observations.
We only analyzed the data for the one beam used to track the target position.  A low-order polynomial baseline subtraction and Hanning smoothing (the velocity resolution is $\sim$ 0.39 \kms\/ after smoothing, reaching a typical rms noise levels of 0.9 Jy) were performed on the averaged spectrum for each target source.

\section{Results} \label{sec:results}
\subsection{KVN Detection } \label{sec:kvn_results} 

\begin{figure}[h]
\mbox{
\begin{minipage}[b]{5.5cm}
\includegraphics[scale=0.3,angle=-90]{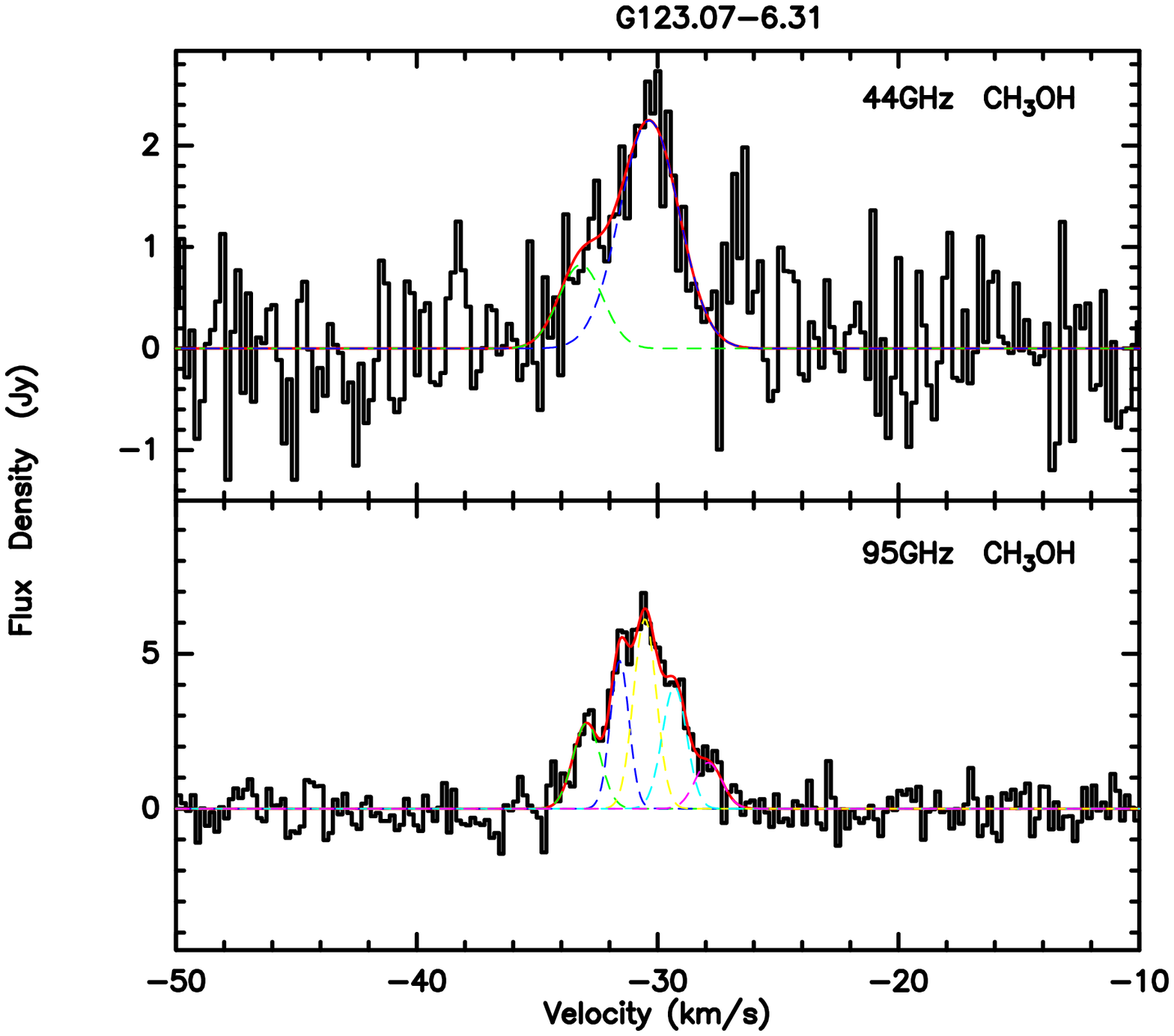}
\end{minipage}
\begin{minipage}[b]{5.5cm}
\includegraphics[scale=0.3,angle=-90]{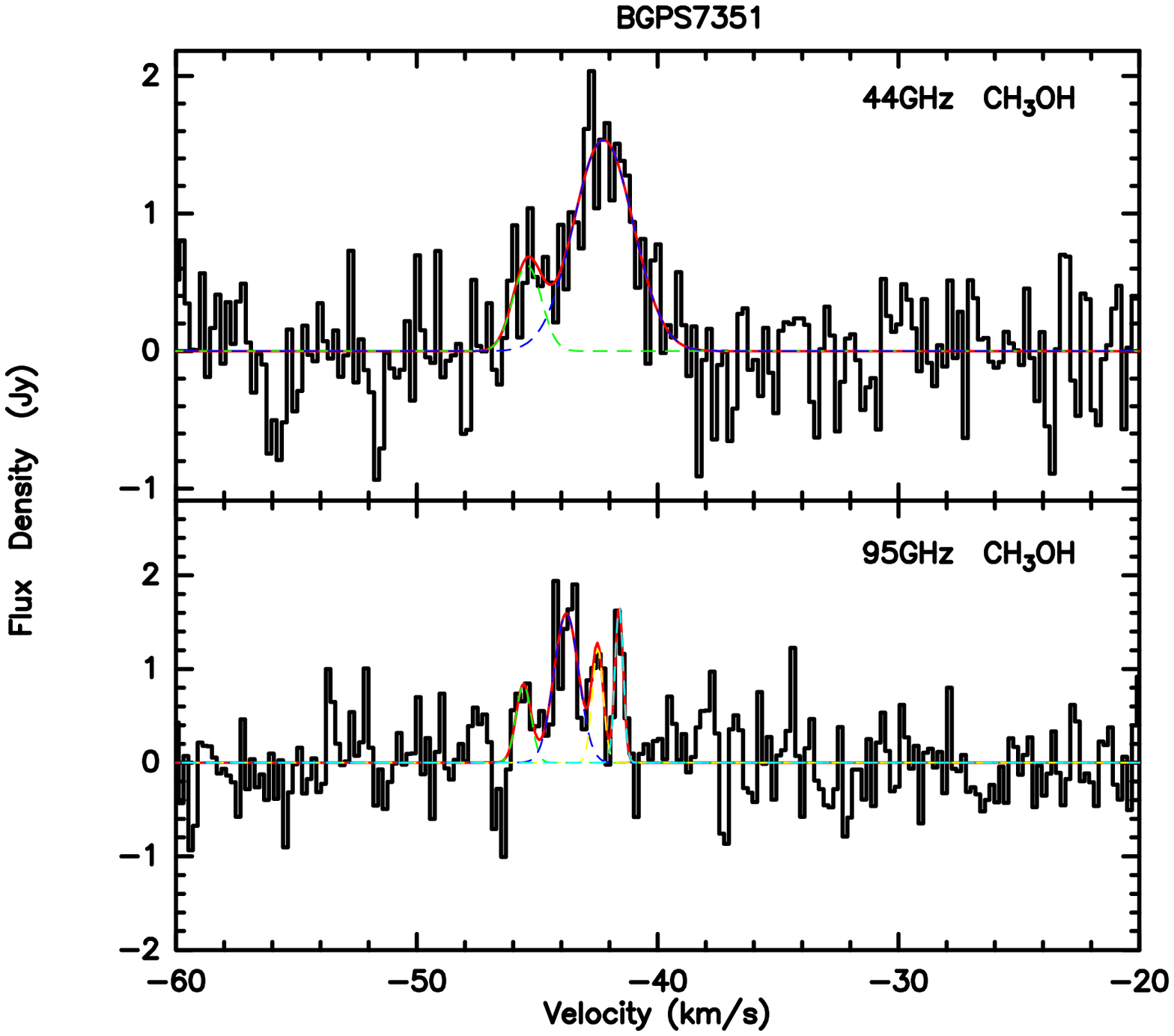}
\end{minipage}
\begin{minipage}[b]{5.5cm}
\includegraphics[scale=0.3,angle=-90]{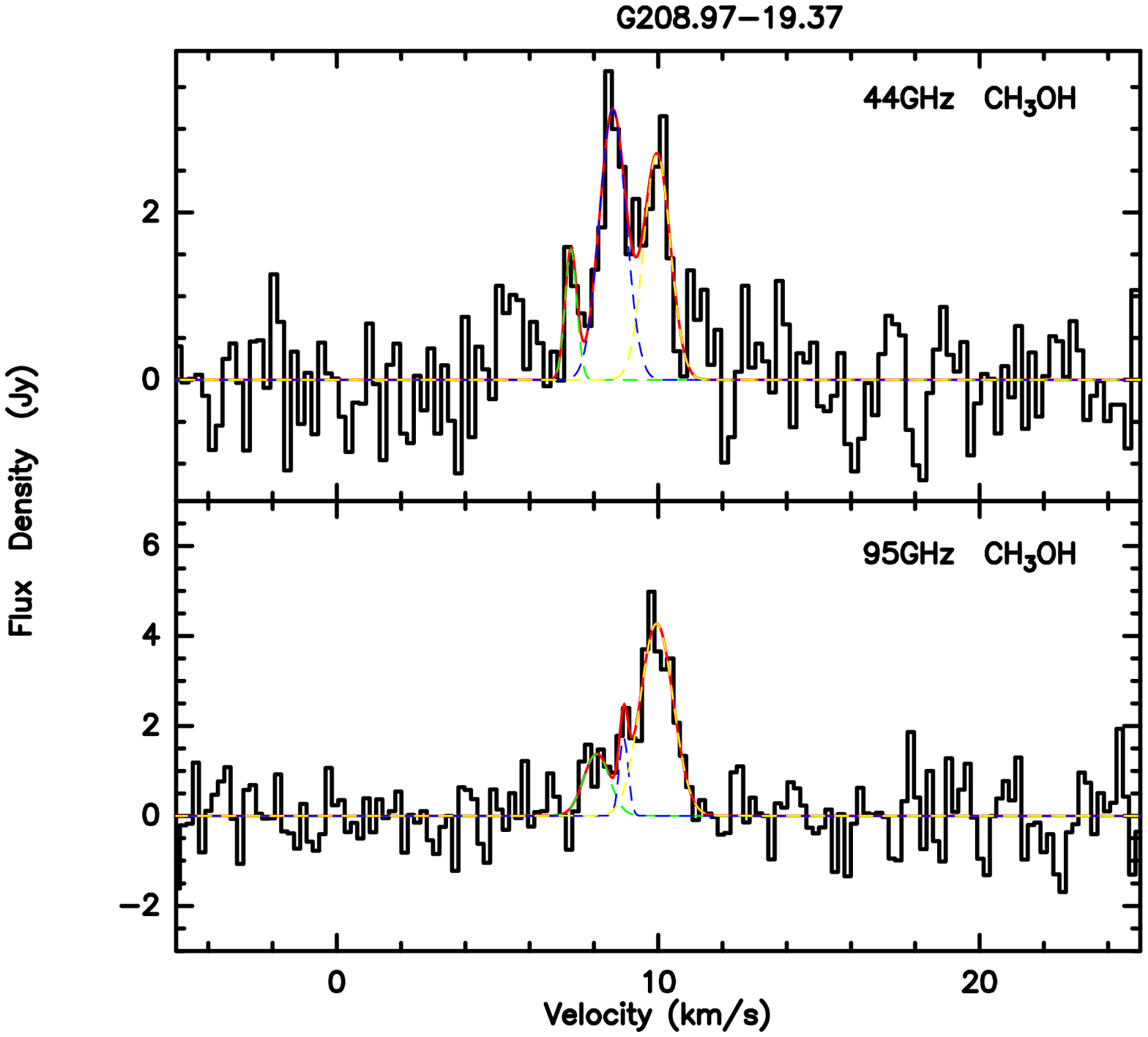}
\end{minipage}
}
\caption{The upper panel and lower panels show the spectra of the 95 GHz and 44 GHz methanol masers, respectively. The bold red line shows the sum of the Gaussian fitting results, while the dashed colored lines show the individual Gaussian fit components.\\
(The complete figure set (144 images) is available in the online material.)\label{fig:spectra}}
\end{figure}

We only consider a source/transition to be a detection where the peak intensity is greater than the 3$\sigma$ noise level in the spectrum.
A total of 144 sites were searched for 44 and 95 GHz methanol masers simultaneously, with 128 (89\%) and 111 (77\%) detections in the 44 and 95 GHz transitions, respectively.  Table \ref{tab:144_obs} summarizes the position of each of the target sources in equatorial coordinates (J2000), whether emission was detected for each transition, whether 44 GHz emission was a new detection, and the noise level for the 44 and 95 GHz spectra, respectively. Tables \ref{tab:44_para} and \ref{tab:95_para} contain the list of all detected sources along with the fitted Gaussian parameters of their spectral features. The spectra of the 144 observed sources are shown in Figure \ref{fig:spectra}.

After cross-matching the latest and largest class I methanol maser catalog \citep{2019AJ....158..233L}, a total of 106 detections at 44 GHz are the first reported observations of maser emission in the target sources.
The exceptions are 17 sources detected in previous KVN observations \citep{2016ApJS..227...17K,2018ApJS..236...31K,2019ApJS..244....2K} and 7 sources with previous interferometric observations (indicated in Table \ref{tab:144_obs}).
One source, G033.390+00.008, which is detected to have weak 44 GHz emission (peak flux density 2 Jy) by \cite{2018ApJS..236...31K}, did not meet the criteria to be considered a detection in the current observations.

The line width of individual 44 GHz emission components obtained from Gaussian fitting ranges from 0.21 to 5.45 \kms\/ with a mean of 1.04 \kms\/ and a median of 0.74 \kms. 
Approximately 80\% (103/128) of the 44 GHz methanol spectra have one or more narrow components (line width $\leqslant$ 1 \kms).
The line width of individual 95 GHz emission components ranges from 0.23 to 9.33 \kms\/ with a mean of 1.07 \kms\/ and a median of 0.82 \kms. 
Approximately 82\% (92/111) of the 95 GHz methanol detections have at least one component with a line width $\leqslant$ 1 \kms.

The better velocity resolution of the current observations shows that nine of the sources at 95 GHz considered maser candidates by \citet[][$\sim$ 0.39 \kms after Hanning smoothing]{2017ApJS..231...20Y}  because they only contained broad spectral components (line width of $>$ 2.5 \kms) do also exhibit narrower spectral components.  With the exception of BGPS2190 (which is better fit with a single broad Gaussian profile), these sources have been fitted with multiple Gaussian components.  
From the current observations there are three sources (BGPS4048, BGPS5057, BGPS7208) at 44 GHz and six sources (BGPS2152, BGPS2190, BGPS4048, BGPS6502, NGC 7538IRAS11, BGPS7208) at 95 GHz that are best fit with only a single broad spectral component.
BGPS7208 (also known as IRAS 23139+5939), has been observed with Very Large Array (VLA), which confirms that although the spectral profile is broad, it is still due to maser emission \citep{2017ApJS..233....4R}.  Single-dish observations alone cannot identify whether these broad emission sources are due to maser or quasi-thermal emission, and future interferometric observations are required to determine this.  For the purposes of our analysis, we have assumed that the detected 44 and 95 GHz  methanol emission in these single broad profile sources is due to maser emission.  In the unlikely event that all of these broader lines are found to be pure thermal emission, the number (fewer than 10) is too small to significantly influence our statistics.

In total, 107 sources have both 44 and 95 GHz emission, 21 sources have only 44 GHz emission, and 4 sources have only 95 GHz emission.
Previous searches have only detected 95 GHz maser emission toward sources which also show 44 GHz masers \citep{2010AA...517A..56F,2018ApJS..236...31K}.
For the four sources where we have a 95 GHz methanol detection without corresponding 44 GHz emission (G183.72$-$3.66, G008.458$-$00.224, BGPS2190, BGPS7318), the 95 GHz emission is weak and only reaches the 3$\sigma$ requirement.  One of the sources, G183.72$-$3.66 (also known as GGD4), was observed at 44 GHz in 2007 March/April using the VLA in the D configuration \citep{2016ApJS..222...18G}, but no emission was detected.  Additional observations of these four sources are required to confirm the 95~GHz detection, and higher resolution observations will be required to determine whether it is maser emission or thermal emission. 

A total of 33 sources were not detected in the 95 GHz methanol emission in the current KVN search.
There are several potential reasons for nondetections:
(1) The majority of nondetected sources were previously observed using the PMO 13.7 m radio telescope, which has a beam size at 95~GHz of $\sim$ 55$\arcsec$.  This is similar to the beam size of the KVN 21m antennas at 44 GHz, but significantly larger than that at 95 GHz, so masing regions detected in previous searches could be located outside the KVN beam at 95 GHz, but within the beam at 44 GHz. There are 21 sources with 44 GHz maser detection but without 95 GHz detection.
(2) For weak masers, the limited integration time means that there is emission with a signal-to-noise ratio lower than three. This emission cannot be unambiguously distinguished from the noise in the spectrum.
(3) The source may have exhibited variability in the 95 GHz class I methanol maser emission.
(4) Some of the sources where a previous detection has been claimed based on data with poor signal-to-noise ratio may have been incorrect, in which case we do not expect to detect either 95 GHz or 44 GHz emission in these sources.

\subsection{Comparison with previous detections}
\subsubsection{95 GHz methanol masers: single-dish observations}  \label{sec:95_compare}

\begin{figure}
\center
\includegraphics[scale=0.8]{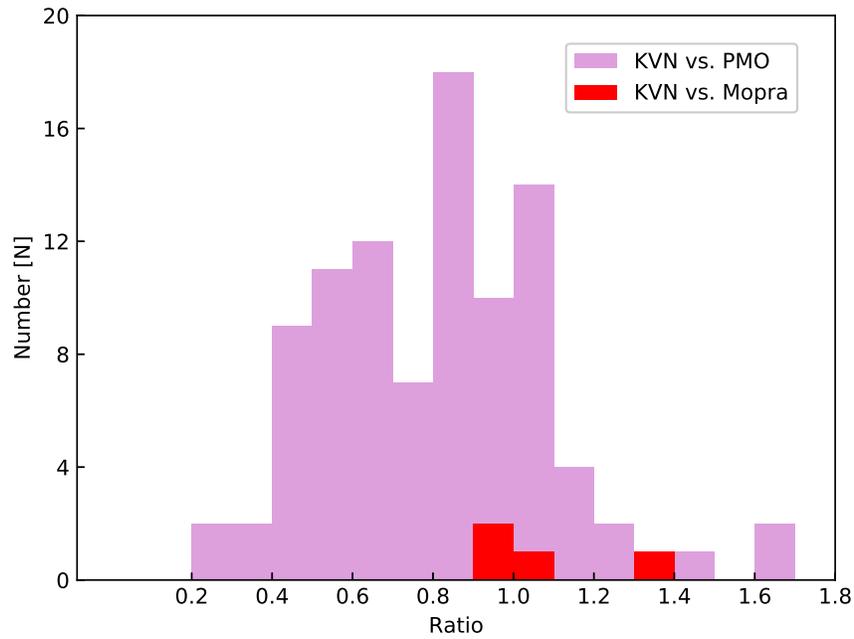}
\caption{The distribution of the peak flux density ratio for the 95 GHz methanol masers detected in the current KVN observations compared to previous PMO (purple) or Mopra (red) observations.\label{fig:ff}\\}
\end{figure}

Excluding the 24 sources with unreliable flux densities due to observations in bad weather conditions (see Section \ref{sec:kvnobs}), there are 99 detected 95 GHz methanol masers, and of these, 95 were previously detected in observations with the PMO 13.7m, while the other 4 were previously detected in observations with the Mopra 22m telescope.  
These 99 sources in KVN observations were smoothed to a similar velocity resolution as the previous detections, i.e., $\sim$ 0.2 \kms\/ for Mopra, and $\sim$ 0.4 \kms\/ for PMO.
Figure \ref{fig:ff} shows the peak flux density ratio of the 95 GHz methanol masers detected in the current KVN observations compared to previous observations with either PMO or Mopra. 
More than 85\% of the sources have a peak flux density ratio between the KVN observations and previous PMO observations ($S_{\rm pk,KVN}$/$S_{\rm pk,PMO}$) that lies in the range 0.4--1.1.
In addition to the flux calibration and pointing accuracy, which affect the ratio, different beam sizes during the observations will lead systematic bias in the ratio.
The beam size of the PMO 13.7m is larger by a factor of two than that of the KVN 21m at 3 mm wavelength.  This means that any maser emission that is offset from the pointing center of the observations will be relatively closer to the edge of the KVN beam than PMO beam.  This will lower the value of $S_{\rm pk,KVN}$/$S_{\rm pk,PMO}$.

The beam size and velocity resolution of the Mopra 22m are $\sim$ 36$\arcsec$ and 0.22 \kms , respectively, which are comparable to those of the KVN 21m. The Mopra observations (2009 August) were made approximately 7 yr prior to the KVN observations (2016 October--December). For the four sources observed with Mopra, the peak flux density ratio is close to one, suggesting no significant variability in the maser emission over that period.

\begin{figure}[h]
\mbox{
\begin{minipage}[b]{4.1cm}
\includegraphics[scale=0.15]{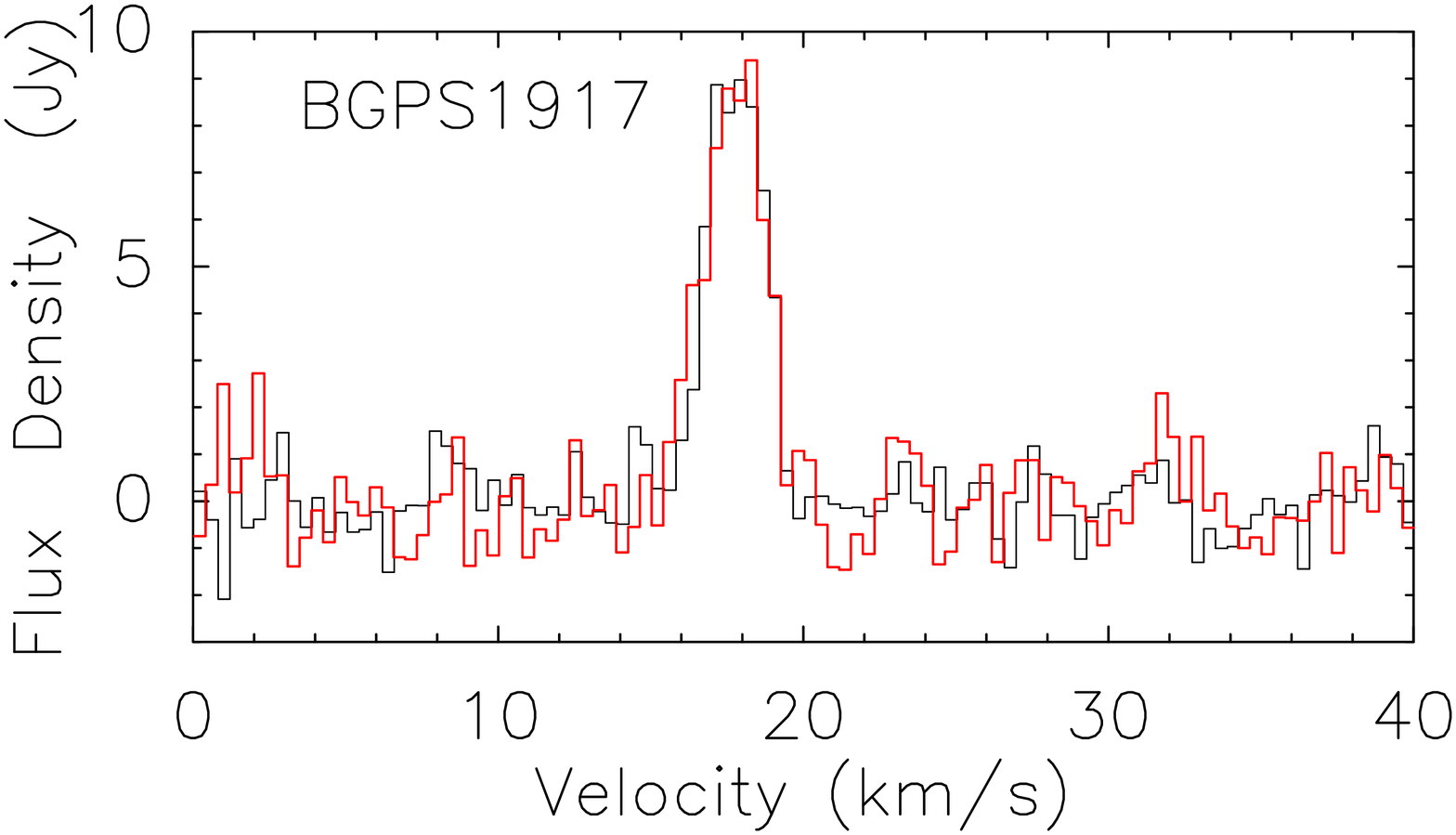}
\end{minipage}
\begin{minipage}[b]{4.1cm}
\includegraphics[scale=0.15]{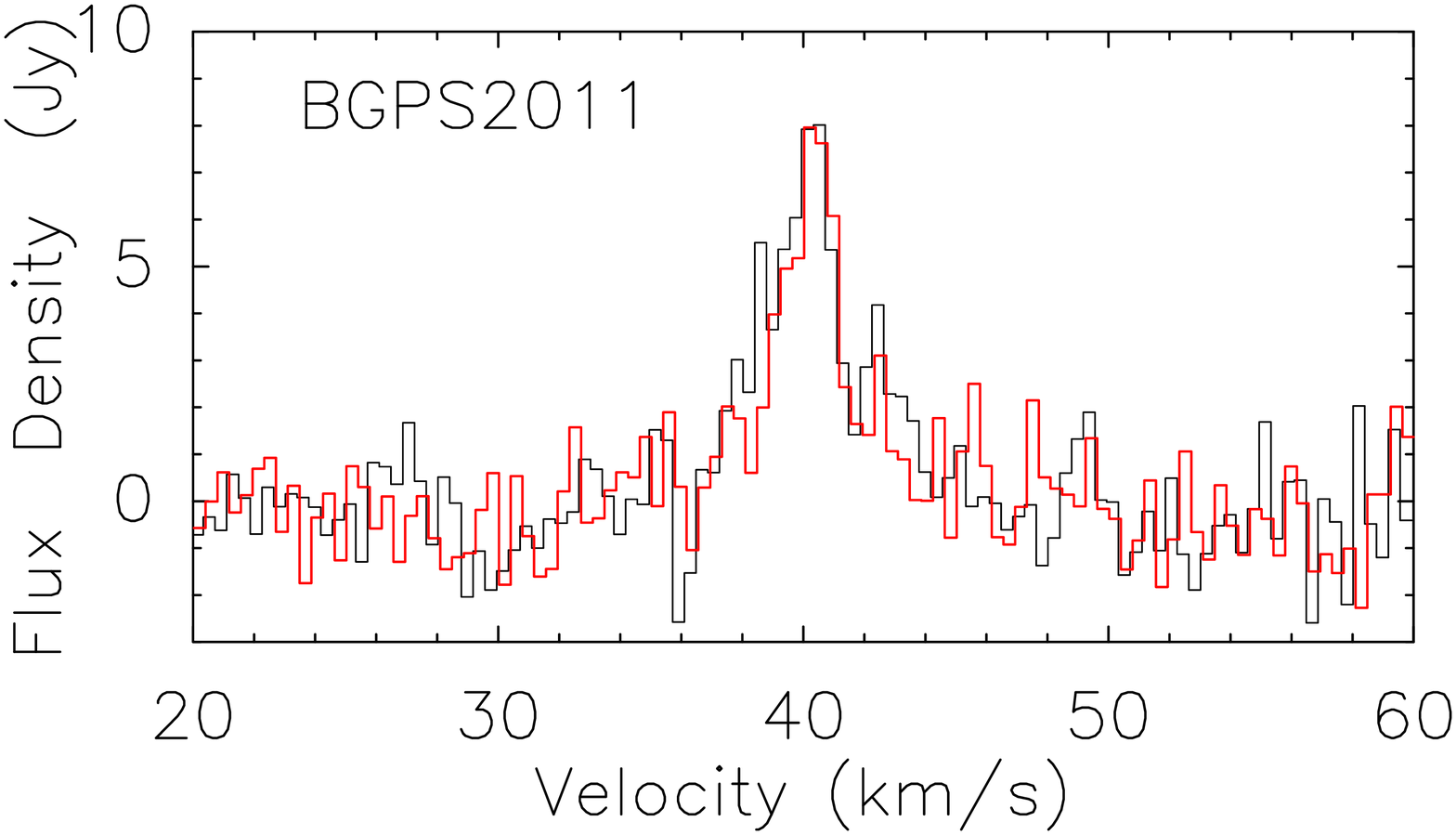}
\end{minipage}
\begin{minipage}[b]{4.1cm}
\includegraphics[scale=0.15]{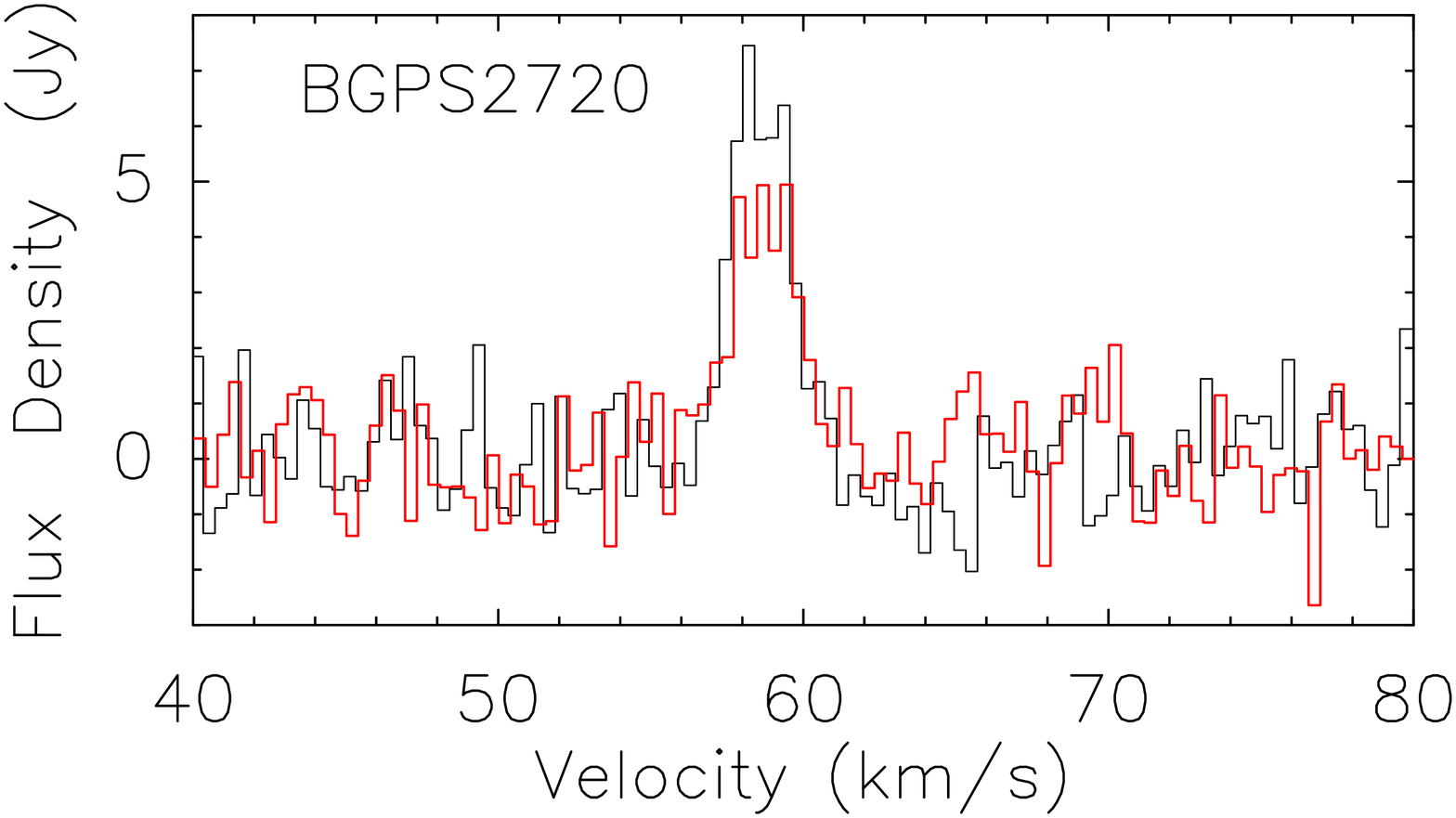}
\end{minipage}
\begin{minipage}[b]{4.1cm}
\includegraphics[scale=0.15]{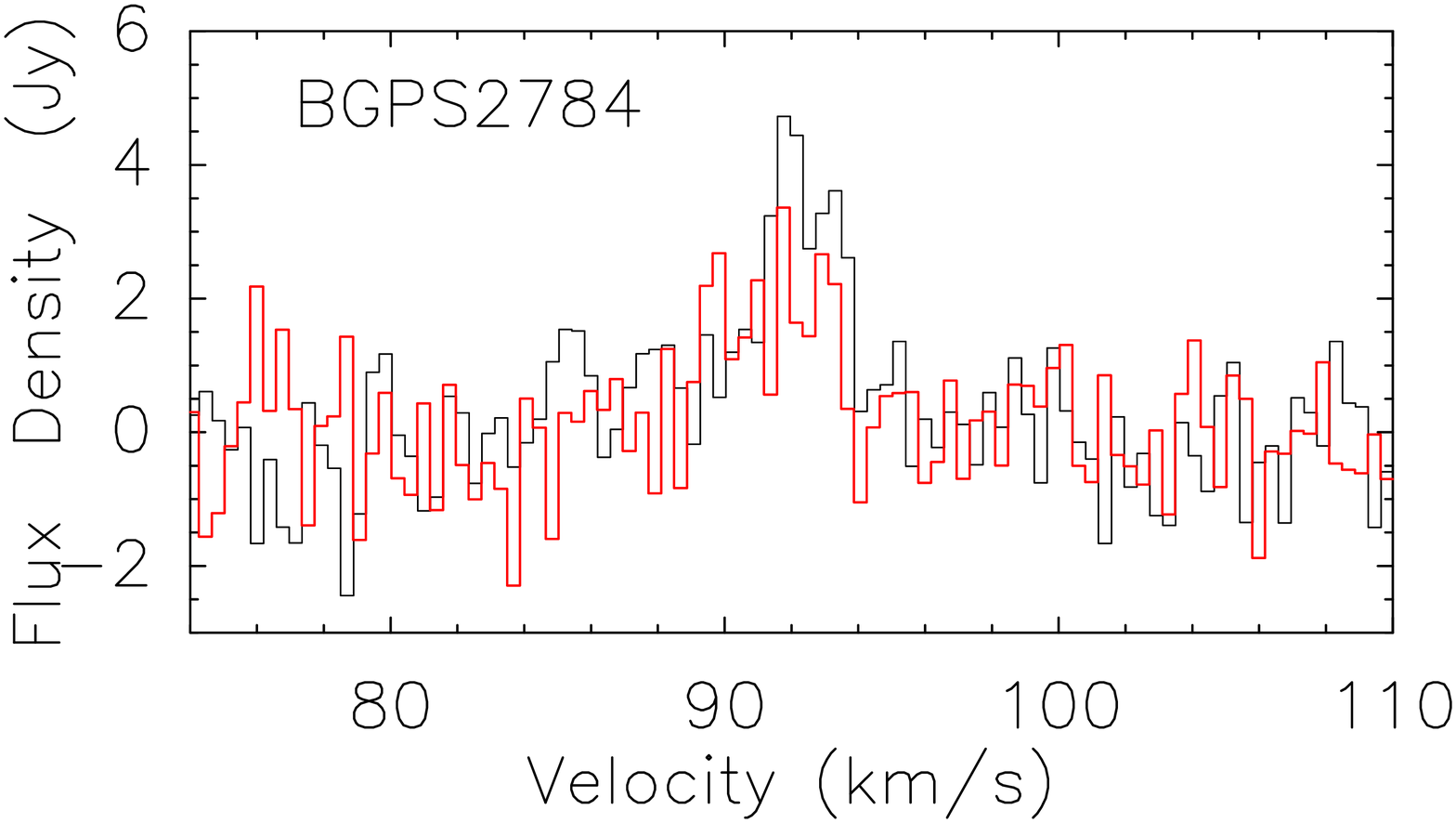}
\end{minipage}
}
\mbox{
\begin{minipage}[b]{4.1cm}
\includegraphics[scale=0.15]{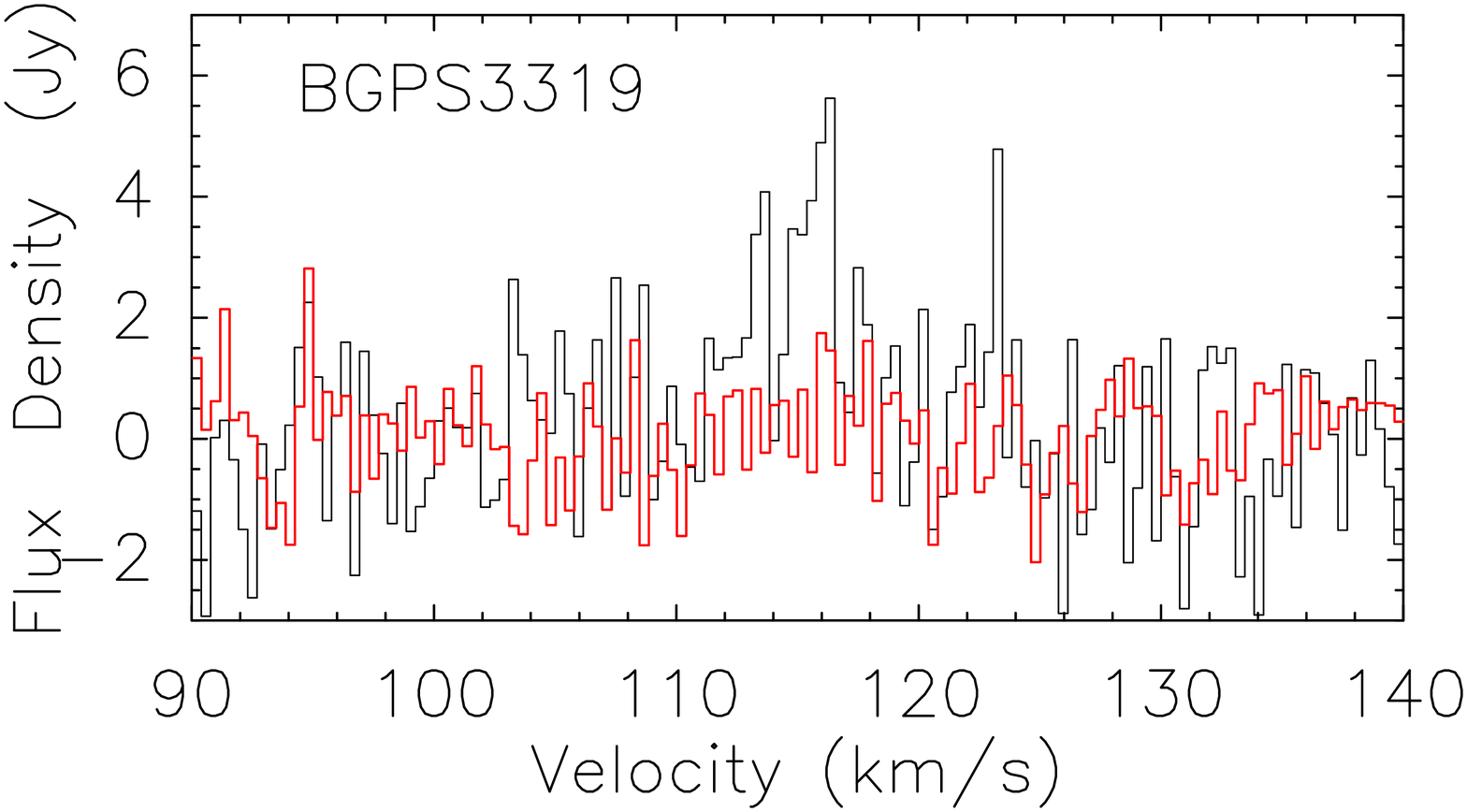}
\end{minipage}
\begin{minipage}[b]{4.1cm}
\includegraphics[scale=0.15]{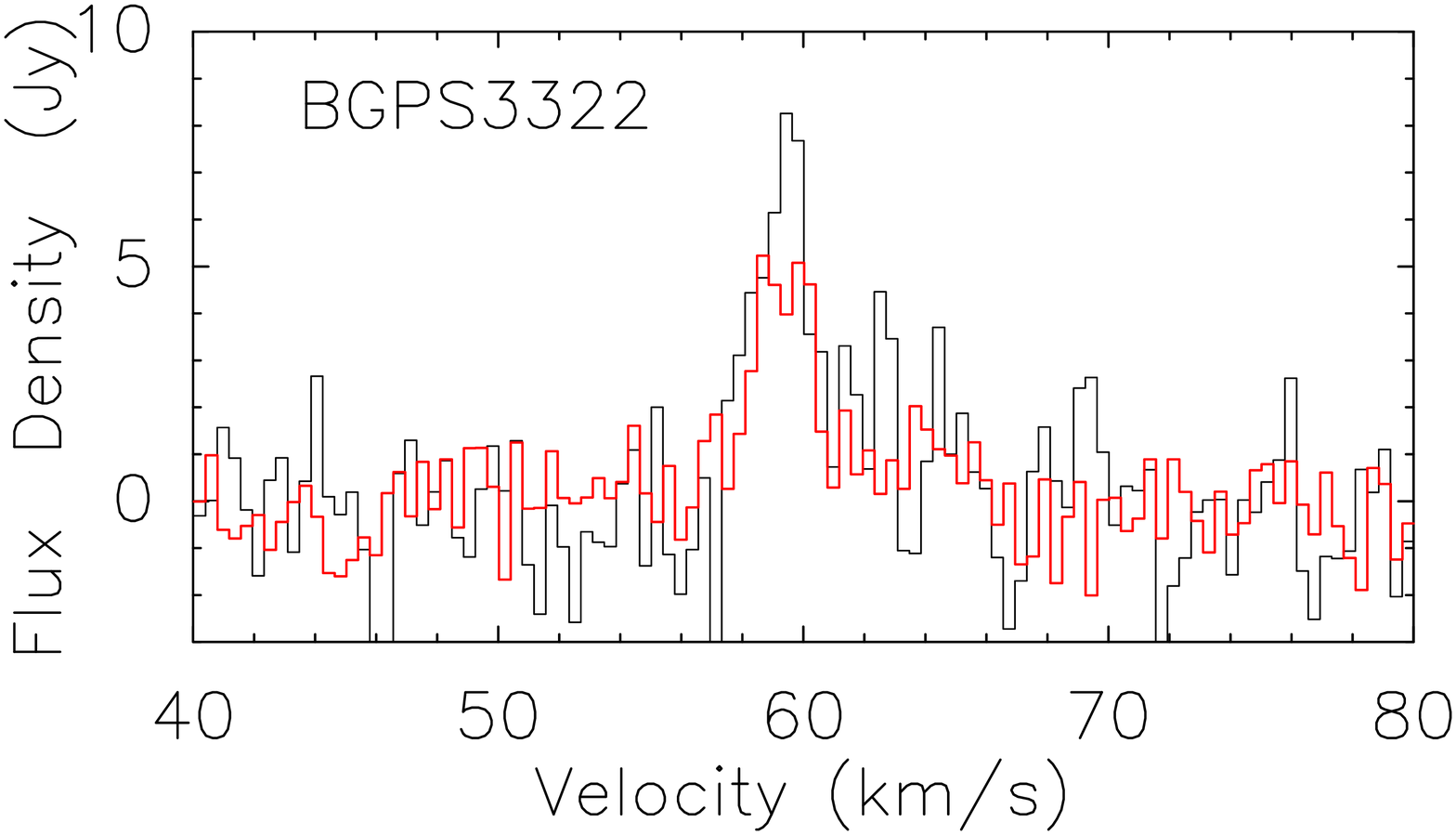}
\end{minipage}
\begin{minipage}[b]{4.1cm}
\includegraphics[scale=0.15]{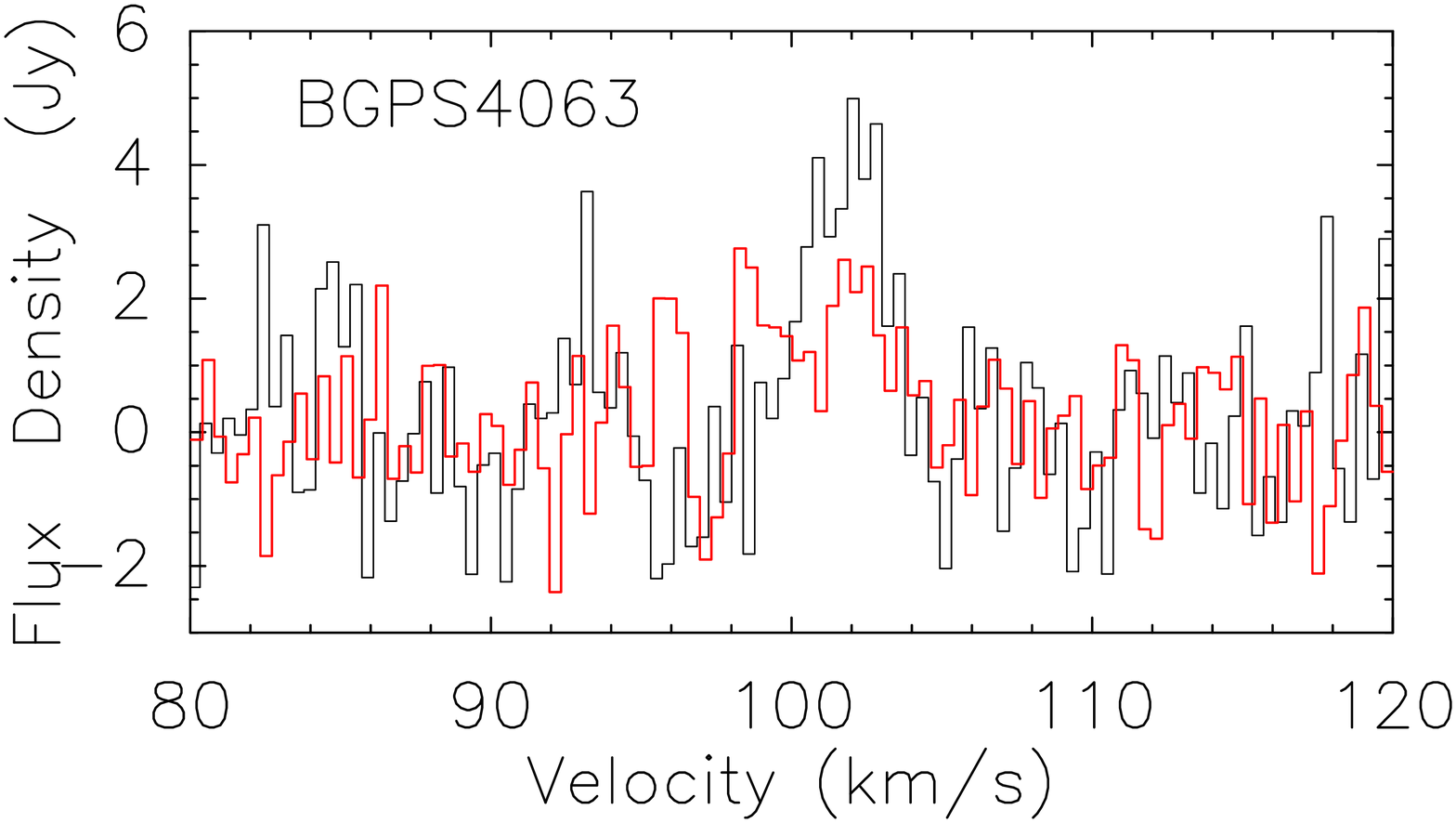}
\end{minipage}
\begin{minipage}[b]{4.1cm}
\includegraphics[scale=0.15]{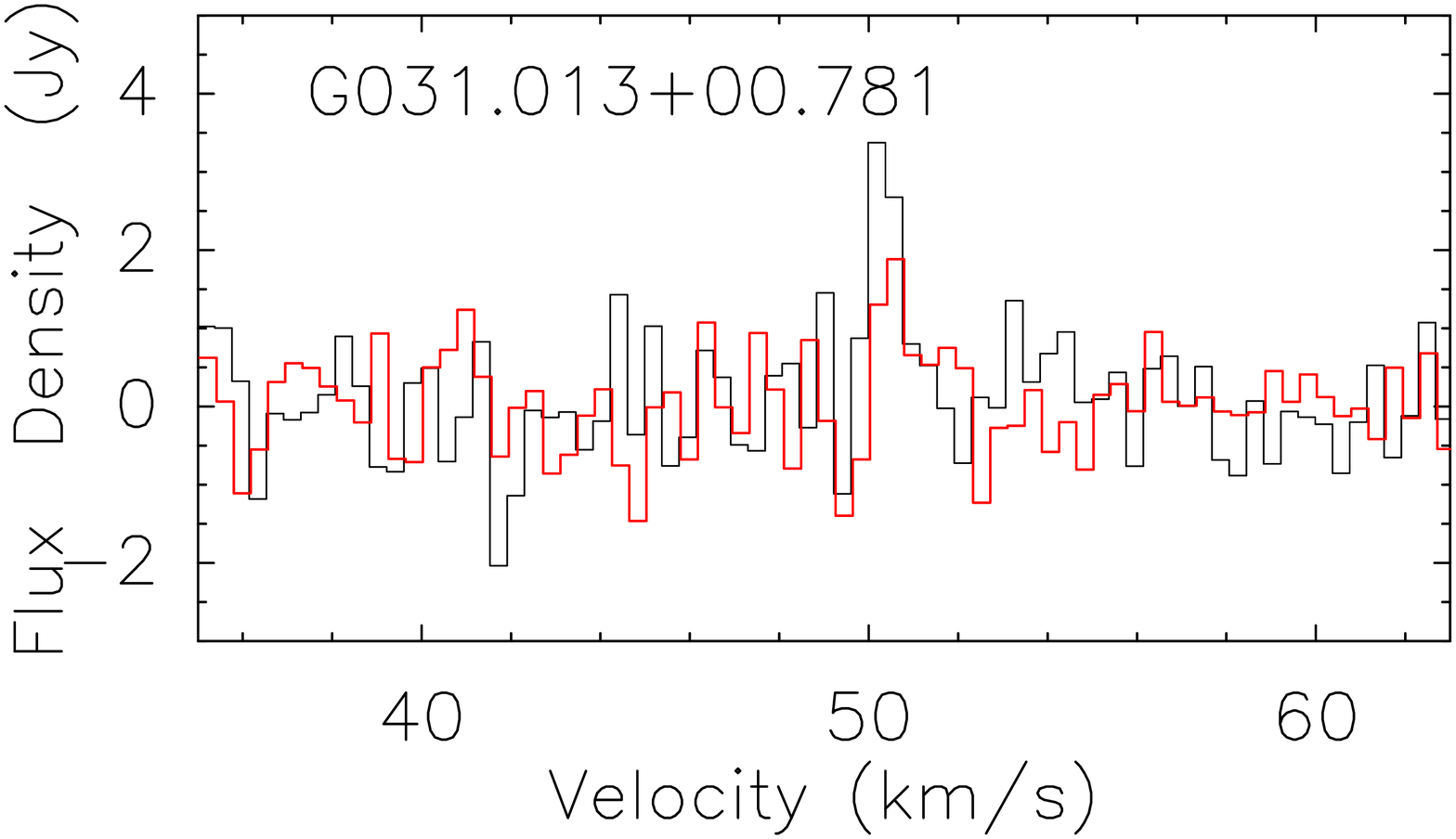}
\end{minipage}
}
\mbox{
\begin{minipage}[b]{4.1cm}
\includegraphics[scale=0.15]{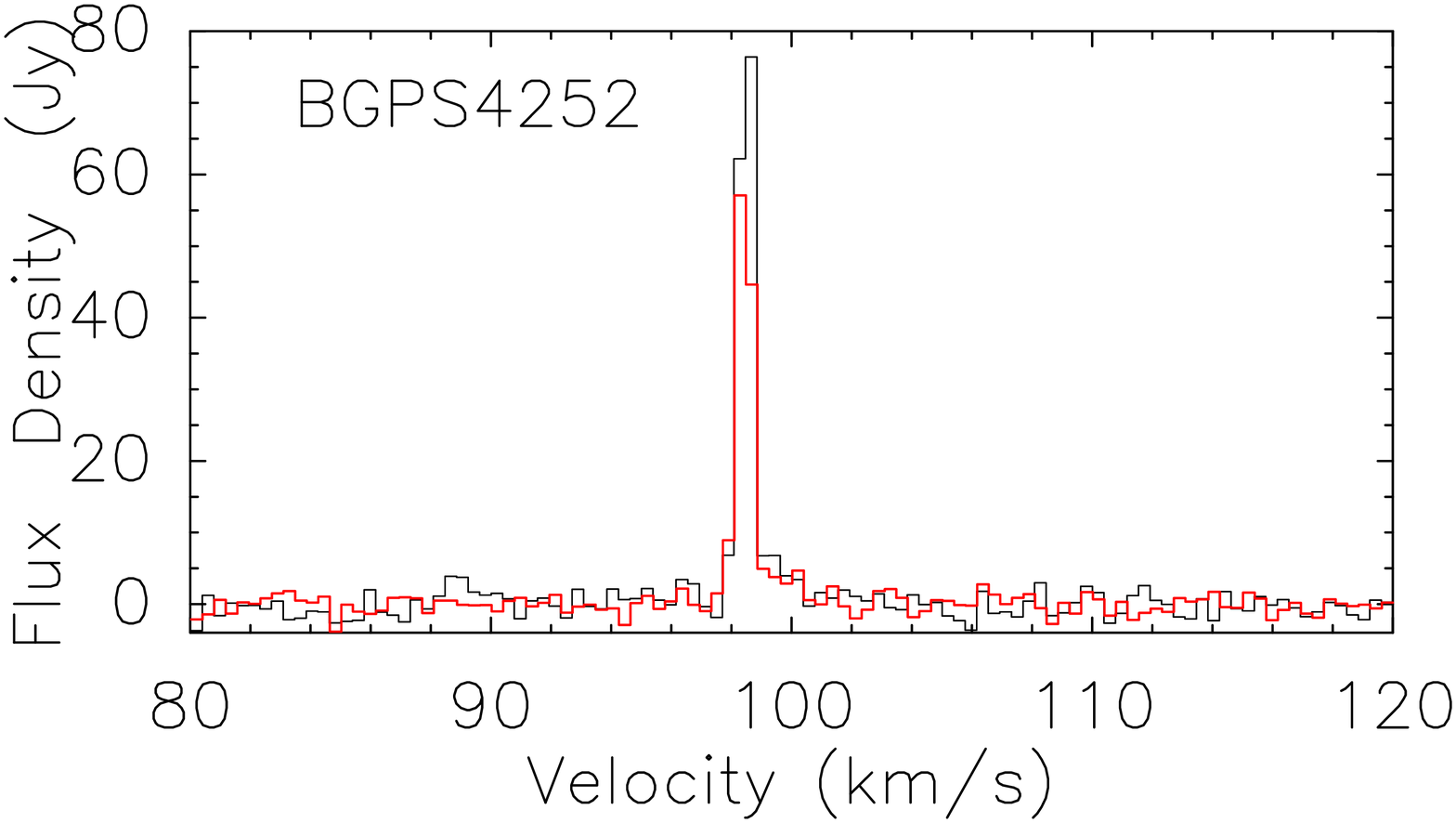}
\end{minipage}
\begin{minipage}[b]{4.1cm}
\includegraphics[scale=0.15]{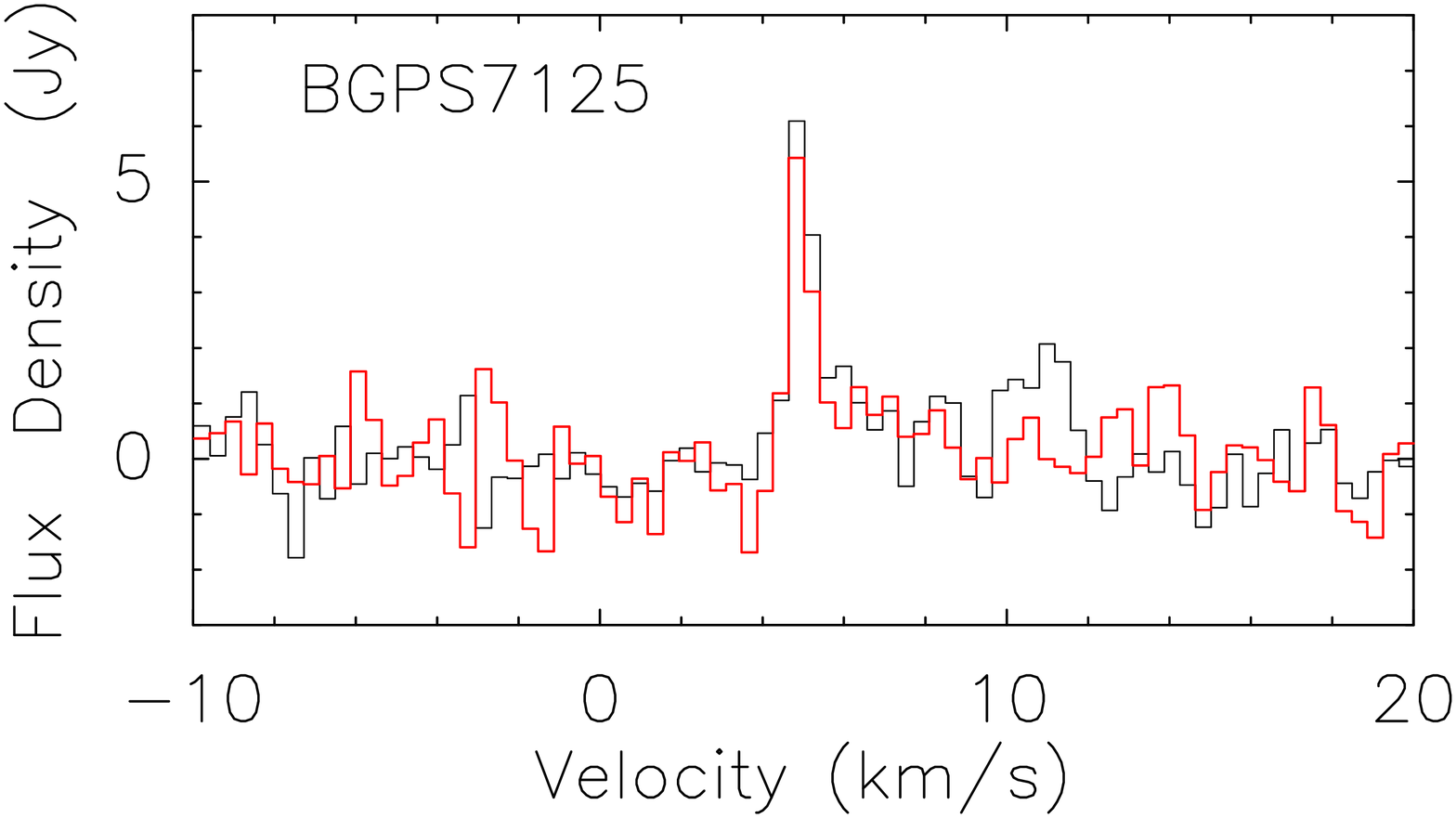}
\end{minipage}
\begin{minipage}[b]{4.1cm}
\includegraphics[scale=0.15]{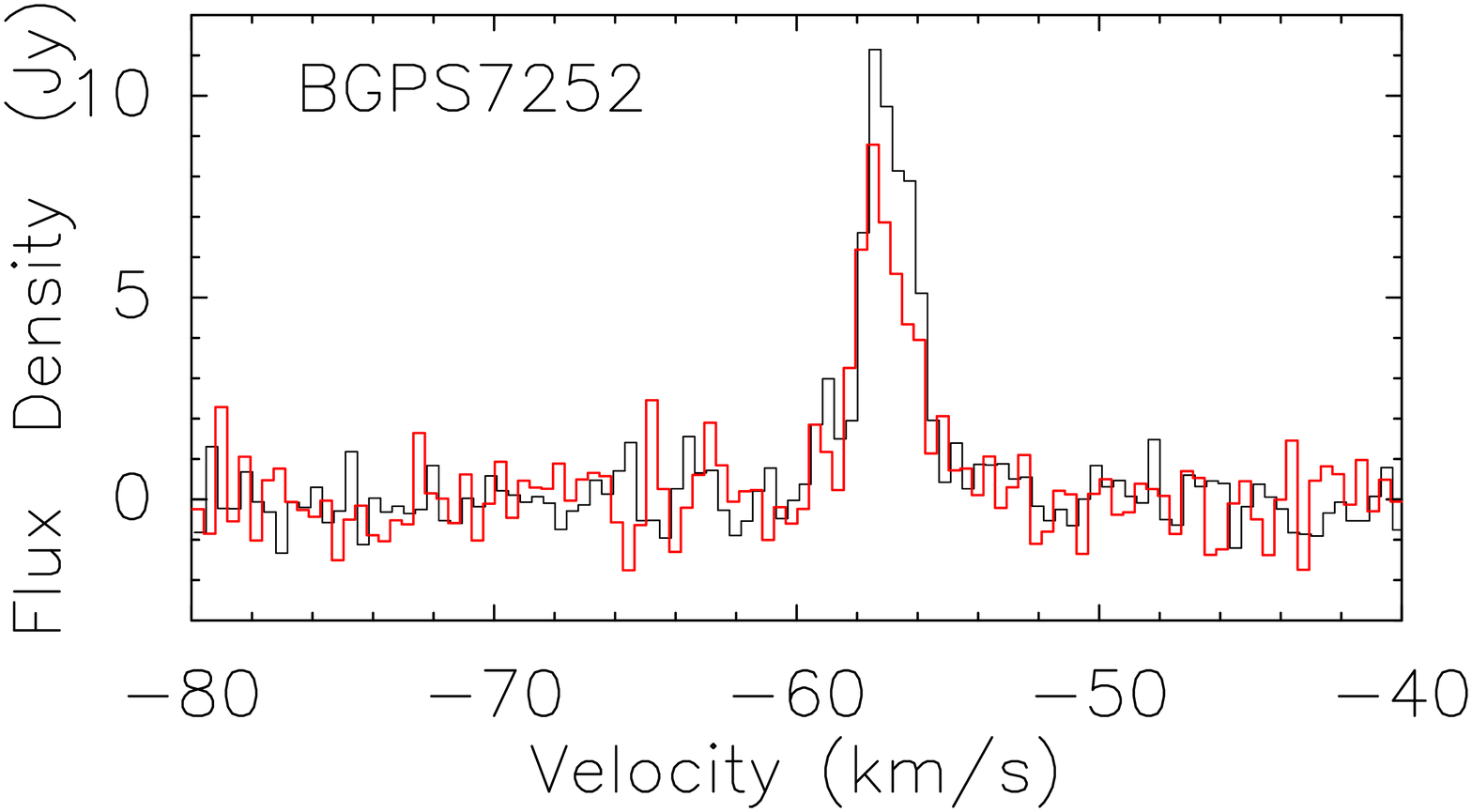}
\end{minipage}
\begin{minipage}[b]{4.1cm}
\includegraphics[scale=0.15]{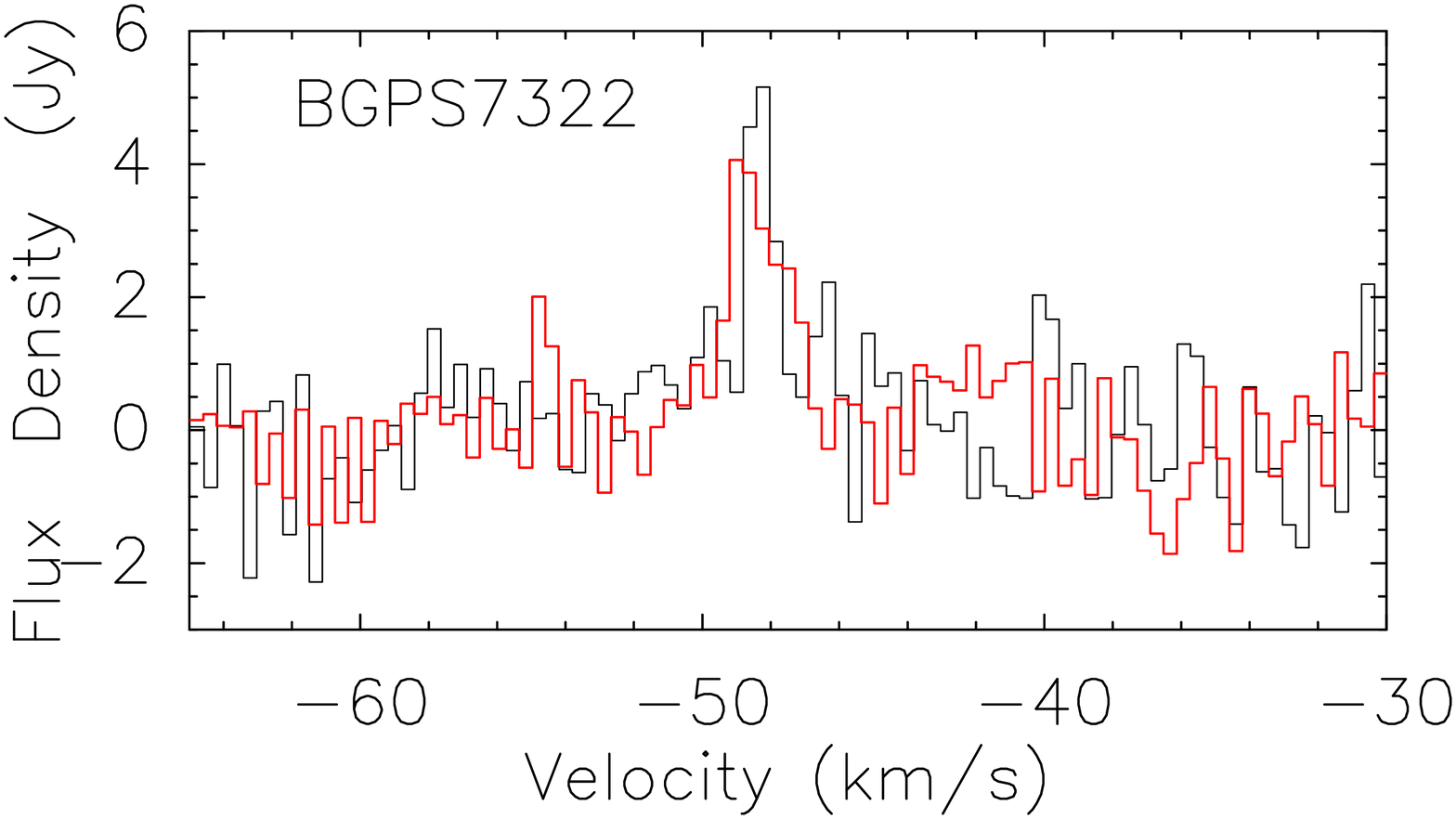}
\end{minipage}
}
\mbox{
\begin{minipage}[b]{4.1cm}
\includegraphics[scale=0.15]{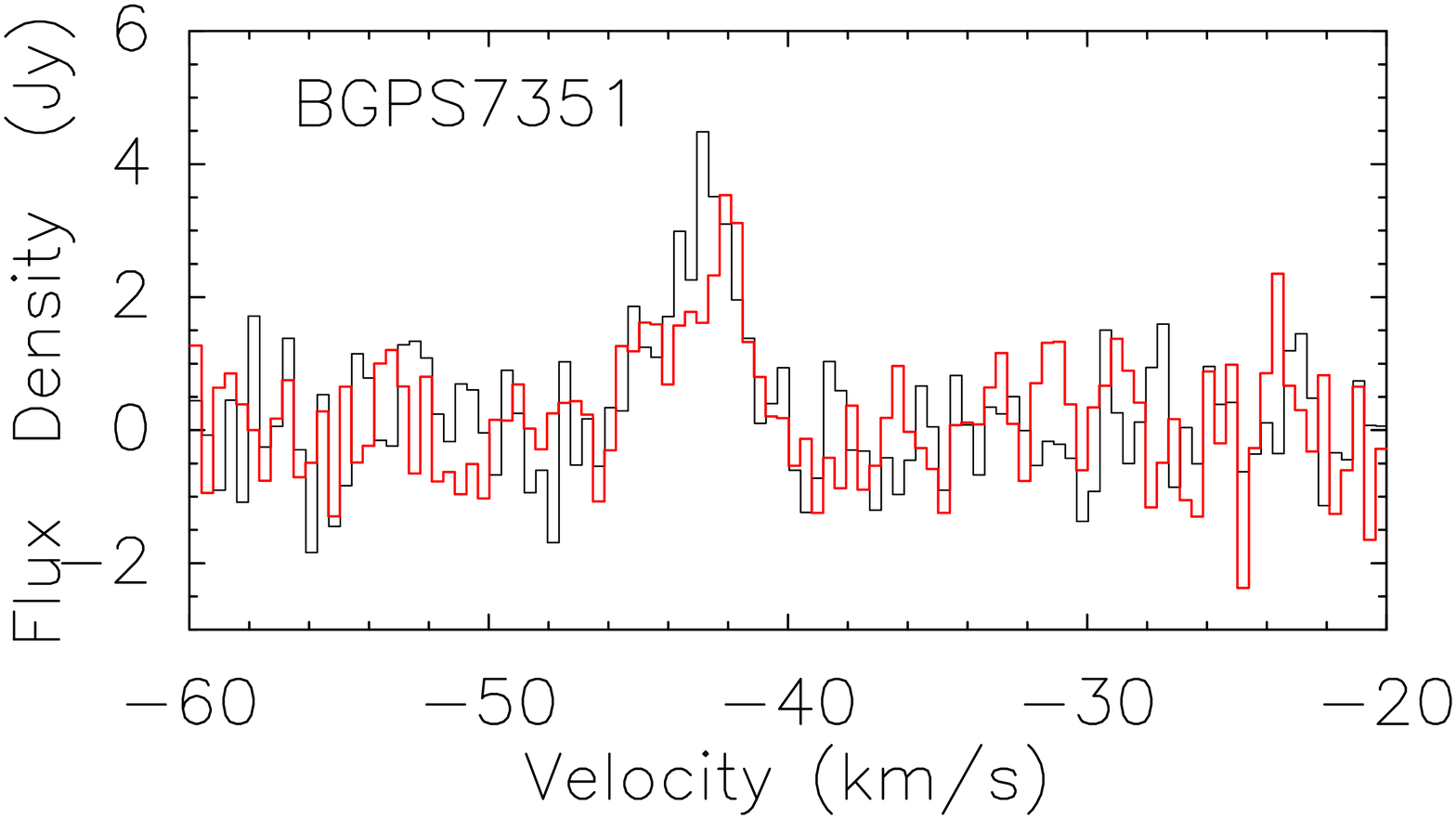}
\end{minipage}
}
\caption{95 GHz methanol emission observed at two different epochs with the PMO 13.7m.  The red spectrum is from the current observations, and the black spectrum shows the previous PMO data.\label{fig:95reobs}\\}
\end{figure}

In total, 13 sources that were previously detected in PMO observations were reobserved using the PMO 13.7m. Nine of them were not detected in the current KVN observations, and four were chosen randomly in the ratio of $S_{\rm pk,KVN}$/$S_{\rm pk,PMO}$.
Table \ref{tab:f_comparison} compares the new PMO observations with the KVN results and previous PMO observations.
Some sources were observed more than once in order to increase the on-source integration time to improve the signal-to-noise ratio. Because there is no evidence of variability in the class~I methanol masers, we averaged all the observed spectra.
Figure \ref{fig:95reobs} shows a comparison of the spectra for each source, with red and black representing the current and previous PMO spectra, respectively.  
Except for BGPS1917 and BGPS2011, whose peak flux densities are consistent between the two epochs of PMO observations, the peak flux densities of other sources are all slightly lower in the current data than before. 

Of the nine sources that are not detected with the KVN at 95 GHz, five were detected in the current PMO 13.7m observations.
This is consistent with the KVN nondetection arising because the class~I methanol maser emission in these sources is offset from the pointing center and lies at the edge of or beyond the narrower KVN beam.  
For a further two sources (BGPS7252 and BGPS7351), the PMO peak flux density is significantly stronger than that of KVN data, and again, this is likely because the maser emission in these sources is offset from the pointing center and toward the edge of the KVN beam.
Two sources (BGPS1917 and BGPS4252) that show a peak flux density in KVN observations that is twice as strong as the PMO detection. When we smooth the KVN spectra to the same velocity resolution as the PMO, the intensity is almost the same, suggesting that these sources have narrow spectral components.

No methanol emission was detected in the current PMO or KVN observations for four sources (BGPS2784, BGPS3319, BGPS4063, and G031.013+00.781).
For BGPS3319 there was no detected 44 GHz methanol emission either, suggesting that this may not be a class~I methanol maser source and that the previously reported detection may not be real, but a misidentification due to a poor signal-to-noise ratio.  
For the other three sources, we detect weak 44 GHz methanol emission, and when we take the typical intensity ratio between 44 and 95~GHz methanol two transitions into account, the 95 GHz emission might be too weak to be detected.  
While we cannot rule out that the uncertainty in the flux density calibration is partly responsible for differences in the measured flux densities between different epochs, an offset between the location of the detected maser emission and the pointing center of the observations appears to be the primary reason for the nondetection of some previously detected 95 GHz methanol masers in the current KVN observations.

To summarize, when we compare the previous 95 GHz detections with new observations of the same transitions made using PMO 13.7 m radio telescope, there is no clear evidence of variability on a timescale of six years (e.g., BGPS3322, BGPS7252, and BGPS7322 in Table \ref{tab:f_comparison}).

\subsubsection{44 GHz methanol masers: interferometric observations}  \label{sec:44_compare}
For nine sources in our sample, 44 GHz interferometric observations from the VLA are available.  These sources are indicated with an asterisk after the name in Table \ref{tab:144_obs}.

\textit{G183.72$-$3.66 (GGD4).}
GGD4 is a low-mass SFR with a known outflow.
This source was observed during 2007 March to April using the VLA in D configuration, but there was no detection of 44~GHz methanol emission \citep[channel rms $\sim$ 52 mJy beam$^{-1}$][]{2016ApJS..222...18G}. In the current observations, we do not detect any emission above the noise at 44 GHz. \cite{2013ApJ...763....2G} reported the detection of a weak 95~GHz class~I methanol maser toward this source, and the current observations also show methanol emission at 95 GHz.
Higher angular resolution observations are needed to determine whether the detected 95~GHz methanol emission is a maser or quasi-thermal.

\textit{BGPS7501 (S255N).} 
\citet{2004ApJS..155..149K} imaged the 44 GHz methanol masers in S255N on 2000 September 25 using the VLA in the D configuration.
The current KVN observations have a similar velocity resolution as those of \citet{2004ApJS..155..149K}, with two major maser features.
The strongest maser component has a velocity of 11~\kms\/ and an intensity of about 250 Jy in the current KVN observations, compared to approximately 150 Jy in the VLA observations.  The secondary maser feature at 10~\kms\/ has a peak flux density of about 60~Jy in both the current and previous observations. 
An interferometric observation resolves out emission structures larger than a specific angular scale that is determined by the shortest baseline and the observing frequency.
\cite{2017MNRAS.471.3915J} analyzed the relative strength of compact maser emission (interferometric measurement) to diffuse emission (simultaneous single-dish measurement), of a sample of 44 GHz class I methanol masers and found that class I methanol emission can be confined to compact structures, extended structures or a combination of the two.
We cannot rule out the possibility that the single-dish KVN spectrum includes additional thermal or diffuse contributions that are resolved out in the interferometric observations, so that only additional high-resolution observations can determine whether this spectral component has varied or not.
In addition to the main spectral features, the VLA observations show a series of weak methanol emission components in the velocity range 7--10 \kms, which are also seen in the KVN spectrum.  
The VLA image shows that the two strongest components are offset from the pointing center of the KVN observations by $\sim$ 11$\arcsec$. 
If the 95 GHz methanol masers are cospatial with the 44 GHz transition, the peak flux density of the 95 GHz methanol masers might be underestimated.
The S255 molecular complex lies between two H {\sc ii} regions, S255 and S257, and is at a distance of 1.59 kpc \citep{2010AA...511A...2R}. 
It consists of three massive star forming regions, S255N, S255IR and S255S. The class I methanol maser emission is located near S255N, whose evolutionary stage is between that of S255IR (more evolved) and S255S (less eveolved). Outflow activity in this region has been studied through a number of different tracers, including CO, shocked H$_2$ and SiO emission \citep{1997ApJ...488..749M,2007AJ....134..346C,2011AA...527A..32W}.

\textit{BGPS3016 (IRAS 18308$-$0841).} 
\cite{2017ApJS..233....4R} reported five 44 GHz methanol maser components in their VLA observations from 2008 September 7.
The three components that are stronger than 6 Jy can be clearly seen in the KVN spectrum and are located approximately 11$\arcsec$ from the pointing center of the KVN beam.
The peak flux density of the 75.8 \kms, 76.4 \kms\/, and 76.9 \kms\/ components measured with the VLA are about 6 Jy, 15 Jy, and 6 Jy, respectively, similar to the current KVN observations, which show 7 Jy, 15 Jy, and 7 Jy, respectively.  Despite the different spatial resolution and flux density calibration uncertainties, we can see that with a similar velocity resolution ($\sim$ 0.16 \kms), there is no evidence for variability of the 44~GHz methanol masers in this source over a period of 8 yr (2008--2016).

\textit{BGPS3307 (IRAS 18337$-$0743).} 
\cite{2017ApJS..233....4R} reported two maser components observed with the VLA on 2008 September 8.
The high-resolution observations show that the class I methanol masers are coincident with the northeastern edge of a H {\sc ii} region traced by bright 8 $\mu$m emission \citep[see figure 1 in][]{2017ApJS..233....4R}.
The H {\sc ii} region, G024.400$-$0.190, has radio recombination line (RRL) emission that peaks at a velocity of 54.7 \kms\/ \citep{2015ApJ...810...42A}, consistent with the methanol maser velocity range (58--62 \kms). 
The H {\sc ii} region is surrounded by a Spitzer dark cloud \citep[SDC;][]{2009AA...505..405P}, G024.405$-$0.187, which is seen in Herschel Hi-GAL data \citep{2016AA...590A..72P}. 
The location of the class I methanol masers suggests that they may be located at the interface between an expanding H {\sc ii} region and an infrared dark cloud (IRDC). On the other hand, \cite{2002ApJ...566..931S} were unable to confirm CO line outflows/wings in this source due to confusion with Galactic plane emission. Higher resolution observations are necessary to identify the possibility of the presence of outflow(s) and further confirm the origin of 44 GHz maser emission.
The two maser components are separated from the KVN pointing center by only 1$\arcsec$, suggesting that the measured flux densities for both the 44 and 95 GHz transitions are reliable.
The peak flux density ratio of the two transitions for both maser components is 0.7, suggesting similar environmental parameters, such as temperature and density.
When we use the kinematic distance estimate of 3.7 kpc obtained from the model of \citet{2014ApJ...783..130R}, the linear separation between the two components is about 0.02 pc. 

\textit{G058.471$+$00.433 (IRAS 19368$+$2239).} 
The KVN observations of this source were made in bad weather, leading to potentially unreliable flux density estimates.
\cite{2016ApJS..222...18G} detected 12 maser components in their VLA imaging of this source. 
The primary and secondary components from the VLA observations are at velocities of 37.9 \kms\/ and 36.1 \kms,  and the separation of these two components from the KVN pointing center are 19$\arcsec$ and 21$\arcsec$, respectively.
These offsets place two of the components beyond the FWHM of the KVN beam at 95 GHz, but strong emission at almost the same velocities was still detected. 

\textit{G65.78$-$2.61 (IRAS 20050$+$2720).} 
\cite{2016ApJS..222...18G} detected a single maser component toward this source with a peak flux density of 1.16 Jy. 
The angular separation of the 44 GHz methanol maser component compared to the KVN beam center is 5$\arcsec$.
Neither 44 nor 95 GHz class I methanol transitions were detected in the current KVN observations (RMS noise level $\sim$0.5 Jy).  Given the low peak flux density of the VLA detection, this is likely because of insufficient sensitivity in the KVN observations.

\textit{BGPS6712 (IRAS 20286$+$4105).} 
\cite{2016ApJS..222...18G} detected two 44 GHz methanol maser components in this source. 
The primary component at a velocity of $-$4.0 \kms\/ has an angular separation of $\sim$ 14$\arcsec$ from the KVN beam center.  The peak flux density measured in the KVN observations (6 Jy) is comparable to that observed by the VLA.  If the 95 GHz methanol maser component at this velocity is cospatial with the 44 GHz counterpart, it lies close to the edge of the KVN beam and the measured flux density will be unreliable. Comparison with previous 95~GHz methanol observations \citep{2017ApJS..231...20Y} shows the peak flux density measured in the KVN spectrum is approximately half of that in previous observations.

\textit{BGPS7208 (IRAS 23139$+$5939).}
\cite{2017ApJS..233....4R} report a single weak 44~GHz maser component with peak flux density of 0.91 Jy  in VLA observations made on 2008 September 12.
The KVN single-dish spectrum shows a peak flux density at 44 GHz above 1 Jy, suggesting that perhaps the maser emission is mixed with thermal/diffuse emission.
The maser component is offset from the KVN beam center by approximately 14$\arcsec$.
Of interest, is that the 95 GHz methanol emission is stronger than the 44 GHz counterpart in this source.

\textit{G111.24$-$1.24 (IRAS 23151$+$5912).} 
\cite{2017ApJS..233....4R} detected two maser components in their VLA observations from 2008 September 12.
For the $-$52.7 \kms\/ component, which is located 5$\arcsec$ from the KVN beam center, the peak flux density detected by both the VLA and KVN is $\sim$4 Jy.
Similarly, for the $-$54.7 \kms\/ component, offset from the KVN beam center by 1$\arcsec$, the peak flux density measured in both single-dish and interferometric observations is 3.5 Jy.  So once again, there is no evidence of variability in this source.
The maser spots are close to the KVN beam center, and hence the measured flux densities of both the 95  and 44 GHz methanol masers will be reliable.
When we assume that the 95 GHz methanol masers are cospatial with 44 GHz transition, the peak flux density ratio between the two transitions measured from the current observations can be considered reliable.
The $S_{\rm pk,95}/S_{\rm pk,44}$ for the $-$52.7 \kms\/ component is about 0.5, while for the $-$54.7 \kms\/ component, it is about 0.9. The difference in the flux density ratio between the two components may reflect a difference in the physical conditions of the environment hosting the two maser components.  The angular separation of the two components is 6.$\arcsec$6, corresponding to a linear separation of about 0.1 pc for a parallax distance of 3.33 kpc \citep{2014ApJ...790...99C}.

\subsection{Emission with single peaks and multiple peaks}

Figure~\ref{fig:spectra} shows a wide variety of spectral profiles for the class I methanol masers. Some show a single, narrow, and strong maser feature, such as BGPS1509, BGPS1584, and BGPS2147, while others (e.g., BGPS3018, BGPS3026, and BGPS3212) contain multiple peaks.  Some sources have broad spectral components, which could be quasi-thermal emission, indicating that the maser emission may in some cases be mixed the thermal emission.
To reliably count the number of maser components, we only considered sources without broad emission components (line width $>$ 2.5 \kms), resulting in 111 and 96 methanol maser sources in our analysis for the 44 and 95 GHz transitions, respectively.  
For the 44 GHz methanol transition, the ratio of the number of sources with a singlep eak to those with multiple peaks is 0.66 (44/67). For the 95 GHz methanol transition, this ratio is 0.57 (35/61).
It is not clear why some sources have a single peak of maser emission while others have multiple peaks, but possible reasons include
(1) precession of the jet/outflow driven by the central object. 
(2) multiple outflows, and
(3) the distribution or structure of the ambient molecular clouds with which the outflow interacts.

\section{Discussion} \label{sec:discuss}
\subsection{The relationship between the two class~I maser transitions}

\subsubsection{V$_{\rm pk}$ of methanol maser features} \label{sec:veldiff}

From Figure \ref{fig:spectra}, it is clear that the 44 and 95 GHz methanol transitions typically cover the same velocity range with very similar radial velocities for individual maser components.  This has been noted in several previous investigations that compared the emission from these two transitions \citep[e.g.,][]{2000MNRAS.317..315V,2018ApJS..236...31K}.   

In our sample, a total of 107 sources have detections for both the 44 and 95 GHz methanol transitions. We have smoothed the spectra from the two transitions to have the same velocity resolution ($\sim$ 0.2 \kms), and the peak velocity of each maser features were determined by Gaussian fitting.
For sources that have the same number of maser features in each transition (e.g., BGPS3307), we matched all of the components individually to analyze the difference in peak velocity.  For sources that have a different number of maser components for the 44 and 95 GHz transitions (e.g., BGPS1917), we compared the velocity of the strongest spectral feature.  

\begin{figure}
\center
\includegraphics[scale=0.5]{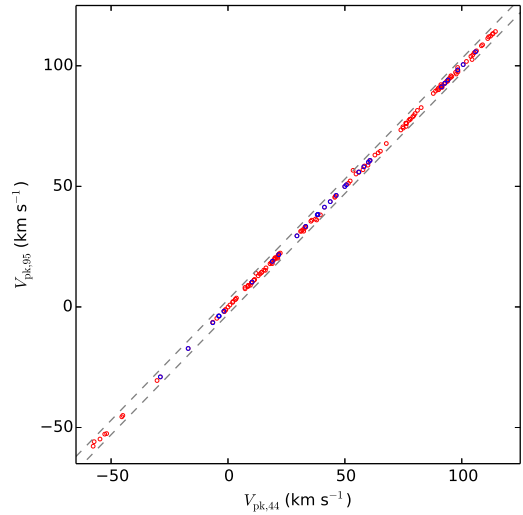}
\caption{Comparison of the velocity of the 44 and 95 GHz methanol peak emission. The two dashed gray lines show a deviation of 3 \kms on both sides. The blue dots represent the 28 sources with only one maser component in both transitions. \label{fig:Vpeak}\\}
\end{figure}

Figure \ref{fig:Vpeak} shows a comparison of the velocity of a total of 161 methanol peak velocities for the 44 and 95 GHz transitions. 
A very strong correlation exists between these two velocities with $V_{\rm pk,95}=(0.9998\pm0.0010)V_{\rm pk,44}-(0.0470\pm0.0614)$, with the Pearson coefficient $r =0.9999$.
For 68\% (109/161) of the maser components, the peak velocity of the two transitions agrees within 0.2  \kms, suggesting that the emission from the 44 and 95 GHz transitions arises in the same environment and is likely cospatial.
The peak velocity difference between the 44 GHz masers and their 95 GHz counterparts, $V_{\rm pk,44}-V_{\rm pk,95}$, ranges from $-$3.15 to 1.37 \kms, with an average of 0.05 \kms and a median of 0.04 \kms. 
BGPS2718 shows the largest velocity deviation between the two transitions, with the strongest emission from the 44 GHz transition at $\sim$54 \kms, but the peak of the 95 GHz transition is at $\sim$57 \kms.  
There is a component at 54 \kms\/ in the 95 GHz spectrum, but it is not the strongest emission from this transition.
The large difference in the radial velocity of the strongest emission between two transitions in this source suggests that there may be significant environmental differences between the two components.
The blue dots in Figure \ref{fig:Vpeak} highlight the 28 sources that have a single maser component for both transitions, which are not subject to the uncertainties caused by fitting multiple Gaussian components. In these sources, the peak emission velocities of two transitions has a tighter correlation, and the deviation ranges from $-$0.23 to 0.2 \kms, with a mean and median for both of $\sim$0.02 \kms.

\subsubsection{Peak flux density}  \label{sec:peak_ratio}

Excluding the sources with unreliable flux densities, the flux density of each component of the 111 44 GHz maser sources was derived from Gaussian fitting of the spectra, ranging from 0.6 to 335 Jy (mean of 12.1 Jy, median 4.7 Jy).  For the 95 GHz masers (99 sources), the flux density range is from 0.6 to 165.4 Jy (mean of 7.4 and median of 3.0 Jy).
BGPS4252 and BGPS7501 are the 44 and 95 GHz methanol masers with the highest detected flux density in our sample, respectively.
There are 95 sources with reliable flux densities in both transitions, and of these, 14 sources ($\sim$ 15\%) have a higher 95 GHz peak flux density than for the 44 GHz methanol maser counterpart.

\begin{figure}
\plottwo{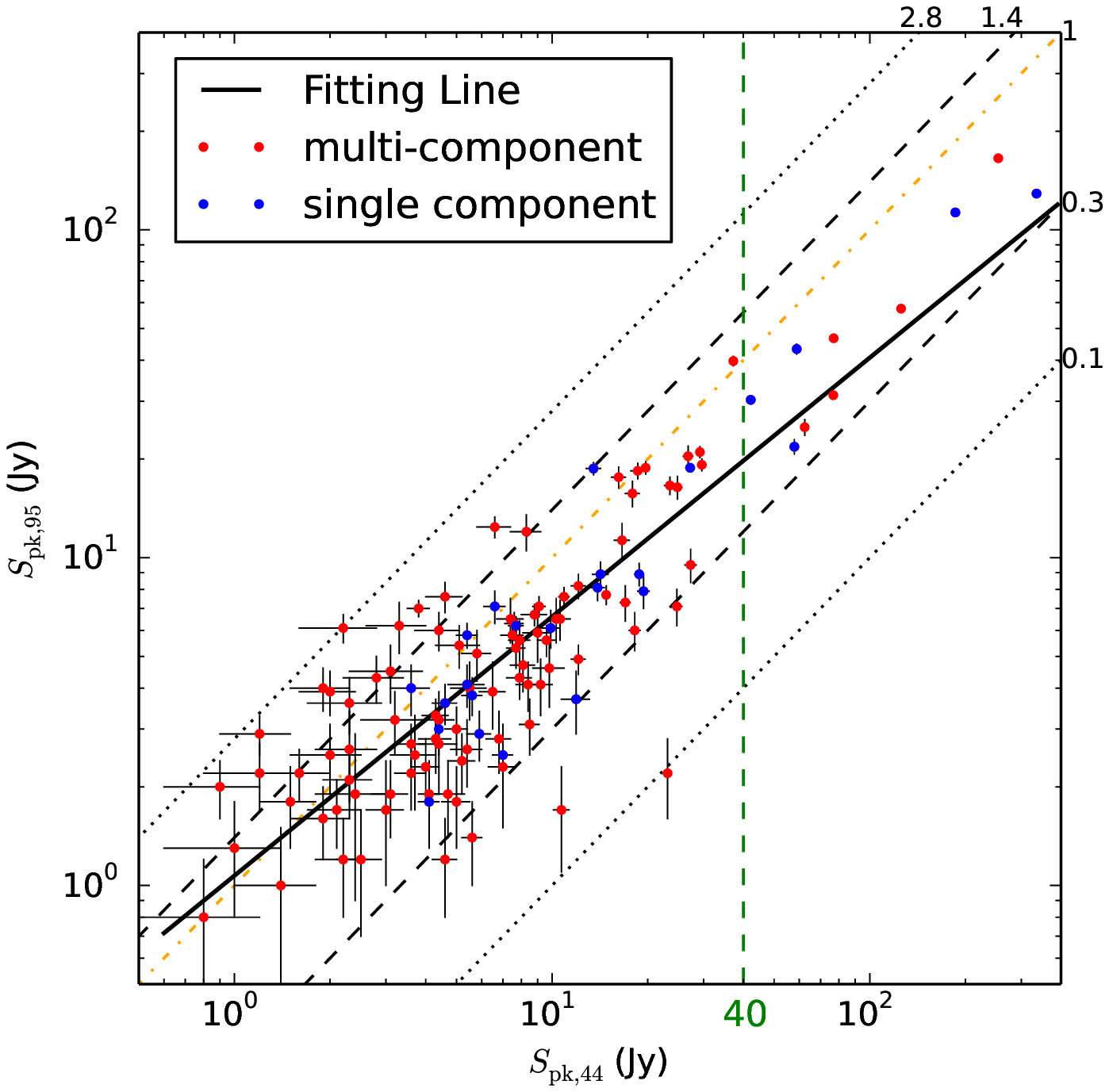}{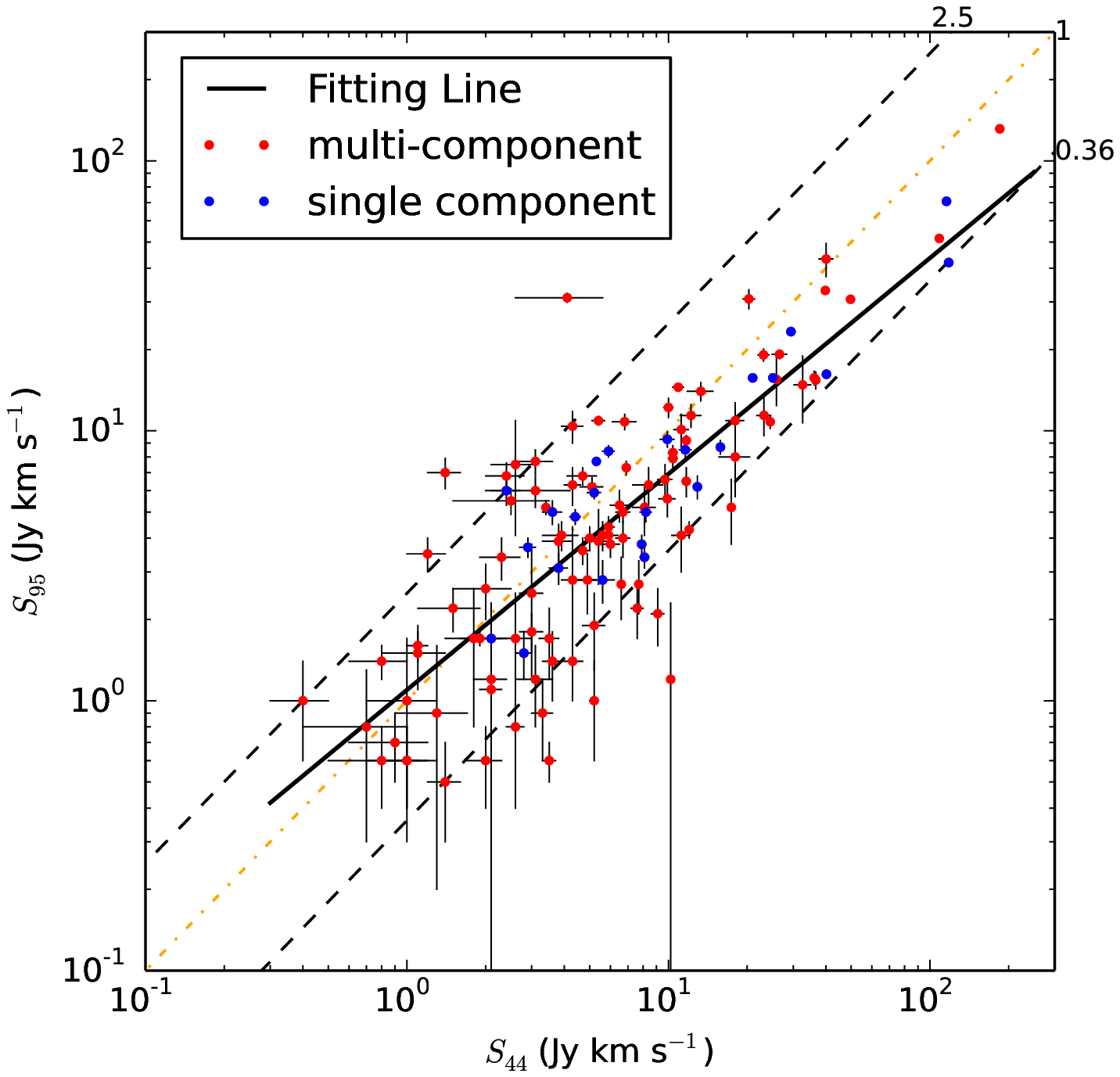}
\caption{Comparison of the peak flux density (left) and integrated flux density of each component (right) between the 44 and 95 GHz methanol emission. The red dots represent the matched methanol emission component in sources with multiple peaks, the blue dots represent the 24 sources with a single methanol emission peak. Error bars are shown in the peak flux density figure. 
The dash-dotted orange line depicts where the two transitions have the same peak flux density, while the solid black line denotes the best linear fit. 
The dotted black lines depict the maximum and minimum value observed for the ratio, the results for the ratio when we consider only single-component sources are denoted by dashed black lines. The dashed green line in the left panel highlights that no source with stronger 95 GHz methanol emissions than 44 GHz emission exists for sources with a 44 GHz peak flux density greater than 40 Jy. \label{fig:flux1}\\}
\end{figure}

Based on the discussion of the velocity difference in Section~\ref{sec:veldiff} and the velocity resolution of the spectra for the two transitions, we can confidently match the emission from the two transitions where there are components with a velocity difference within 0.2 \kms .
Similar to the analysis we have undertaken of the peak velocity, for sources that have the same number of maser features in each transition, we matched all of the components individually to analyze the peak flux density ratio. For sources that have a different number of maser components for the 44 and 95 GHz transitions, we calculated the ratio of the strongest emission in each transition.
In total, 119 pairs were investigated.  Figure \ref{fig:flux1} compares their peak flux density and the corresponding integrated flux density.
The peak flux density ratio, $S_{\rm pk,95}$/$S_{\rm pk,44}$, ranges from 0.1 to 2.8. G123.07$-$6.31 shows the highest ratio, while BGPS4258 has the lowest ratio.

The blue dots in Figure \ref{fig:flux1} highlight the 24 sources with a single maser component in both transitions that are not affected by Gaussian fitting, thus their flux density ratios are more reliable.  There are four sources with a single maser component ($\sim$ 17\%) where the 95 GHz peak flux density surpasses the 44 GHz counterpart.
The ratios, $S_{\rm pk,95}$/$S_{\rm pk,44}$, range from 0.3 to 1.4, with BGPS6547 having the highest value and G8.72$-$0.36 the lowest.
The integrated flux density ratio, $S_{\rm 95}$/$S_{\rm 44}$, is significantly affected by Gaussian fitting, and so we decided to only investigate the range obtained from the sources with a single maser component.  The integrated flux density ratio ranges from 0.36 to 2.5, and this encompasses the ratio for the majority components from the multiple-peak sources.

Figure \ref{fig:flux1} shows strong correlations between the intensity of matched components from the two transitions. Linear fits yield the following results:\\
\begin{equation}
S_{\rm pk,95}=(0.49\pm0.01)S_{\rm pk,44}+(1.88\pm0.66), r=0.96 \\ 
\end{equation}

\begin{equation}
log(S_{\rm pk,95})=(0.79\pm0.04) log(S_{\rm pk,44})+(0.03\pm0.04), r=0.88\\
\end{equation}

\begin{equation}
S_{\rm 95}=(0.58\pm0.01)S_{\rm 44}+(1.55\pm0.59), r=0.93 \\ 
\end{equation}
 
\begin{equation}
log(S_{\rm 95})=(0.80\pm0.05) log(S_{\rm 44})+(0.04\pm0.05), r=0.83\\
\end{equation}

Our linear fit for the peak emission is consistent with that obtained by \citet[][which gives $S_{\rm pk,95}$/$S_{\rm pk,44}$ of 0.56 $\pm$ 0.08]{2018ApJS..236...31K}, and slightly lower than the ratio of 0.71$\pm$0.08 found by \cite{2015ApJS..221....6K}.
When we consider only the 24 sources with a single maser component, the correlation between the intensity of the two transitions becomes stronger, $S_{\rm pk,95}$=(0.43$\pm$0.03)$S_{\rm pk,44}$+(3.79$\pm$2.04), with a Pearson coefficient of 0.96.
A recent investigation using interferometric data from the Australia Telescope Compact Array (ATCA) with an angular resolution of better than 6$\arcsec$ for both transitions found $S_{\rm pk,95}$/$S_{\rm pk,44}$ to be 0.35$\pm$0.10 \citep{2018MNRAS.477..507M}. This result agrees with earlier estimates of 0.31 and 0.32 \citep{2000MNRAS.317..315V,2015MNRAS.448.2344J}.
However, for the three studies that have obtained ratios of about 0.31--0.35, these ratios were not obtained under the same conditions, such as the same velocity resolution, observing mode, and epoch.
The three studies all used data from different epochs. Moreover, \cite{2015MNRAS.448.2344J} compared their 44 GHz maser fluxes from ATCA observations with 95 GHz fluxes from previous single-dish studies, and \cite{2018MNRAS.477..507M} compared the two data sets at very different velocity resolutions.
The current KVN observations observed the two transitions simultaneously with the same pointing, but for the majority of sources, we do not have interferometric observations, and so the position of the maser components within the beam is unknown. 
We cannot rule out the possibility that some masers are close to the edge of the KVN beam at 95 GHz, but this will cause the detected flux density to be lower than the real value and hence lead to a peak flux density ratio of 95 and 44 GHz transitions that underestimates the real situation.

Class I maser pumping models \citep[e.g.,][]{2016A&A...592A..31L,2018IAUS..336...57S} find that the line ratio between maser transitions is sensitive to the physical conditions of the environment. 
The different line ratios between the two transitions found in previous studies may be due to differences in the sample selection criteria. In addition, interferometric observations \citep{2017MNRAS.471.3915J} have found class I methanol emission can consist of only compact components, extended components, or a combination of the two. 
We cannot rule out that the single-dish spectra are a blend of maser and quasi-thermal emission in some sources, which may cause line ratio differences between interferometric and single-dish observations.
The class I methanol masers reported in \cite{2015ApJS..221....6K} and \cite{2018MNRAS.477..507M} all  have an associated 6.7 GHz methanol maser, but the former were observed by the KVN in single-dish mode, while the latter were observed using ATCA, and the line ratio between the 95 and 44 GHz methanol transitions measured from the two studies is quite different.

\begin{figure}
\center
\includegraphics[scale=0.8]{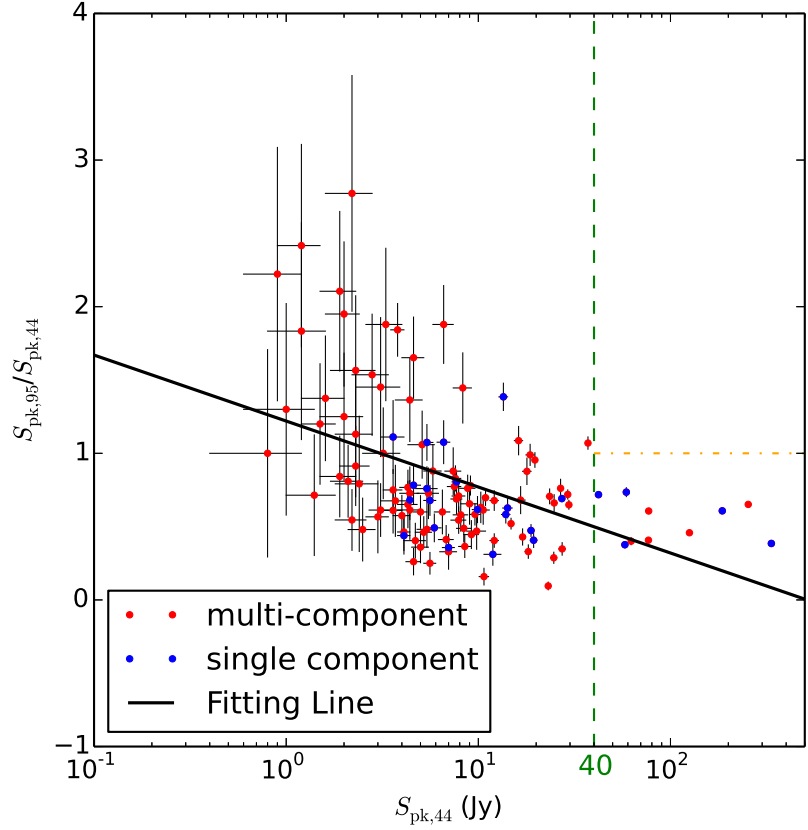}
\caption{The peak flux density ratio of $S_{\rm pk,95}$/$S_{\rm pk,44}$ against with the peak flux density of the 44 GHz methanol maser. As shown before, the red dots represent the matched methanol emission component in sources with multiple peaks, and the blue dots represent the 24 sources with a single methanol emission peak. Error bars are shown in the peak flux density figure. The dash-doted orange line depicts where $S_{\rm pk,95}$/$S_{\rm pk,44}$ equals 1 when the peak flux density of 44 GHz is stronger than 40 Jy (dashed green line). The solid black line denotes the best linear fit. 
\label{fig:ratio}\\}
\end{figure}

It should be noted that $S_{\rm pk,95}$/$S_{\rm pk,44}$ noticeably decreases as the flux density of the 44 GHz transition increases. Figure \ref{fig:ratio} shows $S_{\rm pk,95}$/$S_{\rm pk,44}$ against the peak flux density of 44 GHz methanol emission. The best linear fit is 
$S_{\rm pk,95}$/$S_{\rm pk,44}$=($-$0.45 $\pm$ 0.08)$log(S_{\rm pk,44})$ + (1.22 $\pm$ 0.08), with a Pearson coefficient of $-$0.47. 
As the 44 GHz peak flux density increases, the $S_{\rm pk,95}$/$S_{\rm pk,44}$ decreases and flattens to an approximately constant value lower than one.
From both Fig. \ref{fig:flux1} (left panel) and Fig. \ref{fig:ratio}, it is clear that when the peak flux density of the 44 GHz methanol maser is lower, the ratio of $S_{\rm pk,95}$/$S_{\rm pk,44}$ has more scatter. 
In addition, for the 10 components (from 119 methanol transition pairs) where the peak flux density of the 44 GHz transition is greater than $\sim$ 40 Jy, there is no case where the 95 GHz methanol emission exceeds that of the 44 GHz counterpart.
A possible explanation for this behavior is that the sensitivity of the 44 and 95 GHz transitions to the physical parameters is different, e.g., the methanol specific column density \cite[see the definition in][]{1994A&A...291..569S,2005MNRAS.360..533C}, which is a factor in the maser optical depth of the maser and therefore determines the maser brightness. Pumping models for class I methanol masers (A.M. Sobolev 2020, private communication, and S.Yu. Parfenov 2020, private communication) show that the 95 GHz transition becomes relatively weaker than the 44 GHz transition with the increase in maser flux. This dependence is not linear and is more pronounced when the masers have lower fluxes.

\subsection{Masers and Central Objects}   \label{sec:iras}

Because the angular resolution of the current single-dish observations is limited, we cannot study the infrared emission where the maser arises in depth, but only the overall infrared environment of the central YSOs that host the maser.

The Infrared Astronomical Satellite (IRAS\footnote{\url{http://irsa.ipac.caltech.edu/Missions/iras.html}}) performed an unbiased sky survey at 12, 25, 60, and 100 $\mu$m, with an angular resolution ranging from 30$\arcsec$ to 2$\arcmin$.  The data from this survey were used to create the IRAS Point Source Catalog (version 2.1) which we have used to analyze the environment of the maser host.
Of the 128 detected 44 GHz methanol masers, 59 have an IRAS counterpart within 1$\arcmin$, and 
Table \ref{tab:iras} gives the angular separation and the name of the \textit{IRAS} counterpart of each source, as well as the \textit{IRAS} band fluxes, distance, and bolometric luminosity.
We used $F_{x}$ to denote the flux density in the $x$ $\mu$m IRAS band in Jy, and $Q_{x}$ to denote the corresponding flux quality, with values of 1, 2, and 3 representing an upper limit, moderate quality, and high quality, respectively.
We also calculated the bolometric luminosity of the central object based on the IRAS fluxes \citep{2007AJ....133.1528C}.
Distances for each of the target sources were tabulated by \cite{2017ApJS..231...20Y}, primarily using kinematic distances calculated from the \cite{2014ApJ...783..130R} model.

\begin{figure}
\center
\includegraphics[scale=0.4]{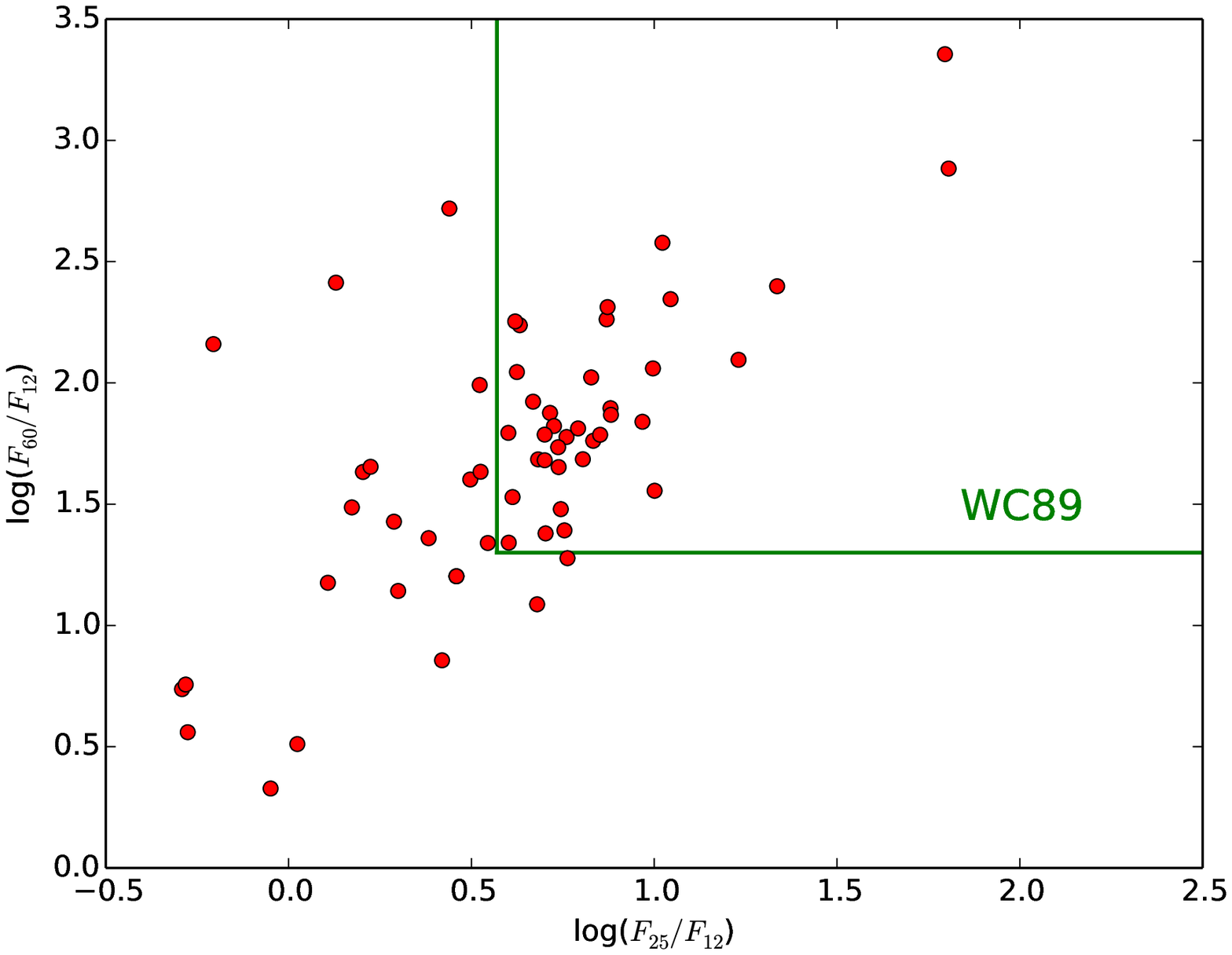}
\includegraphics[scale=0.4]{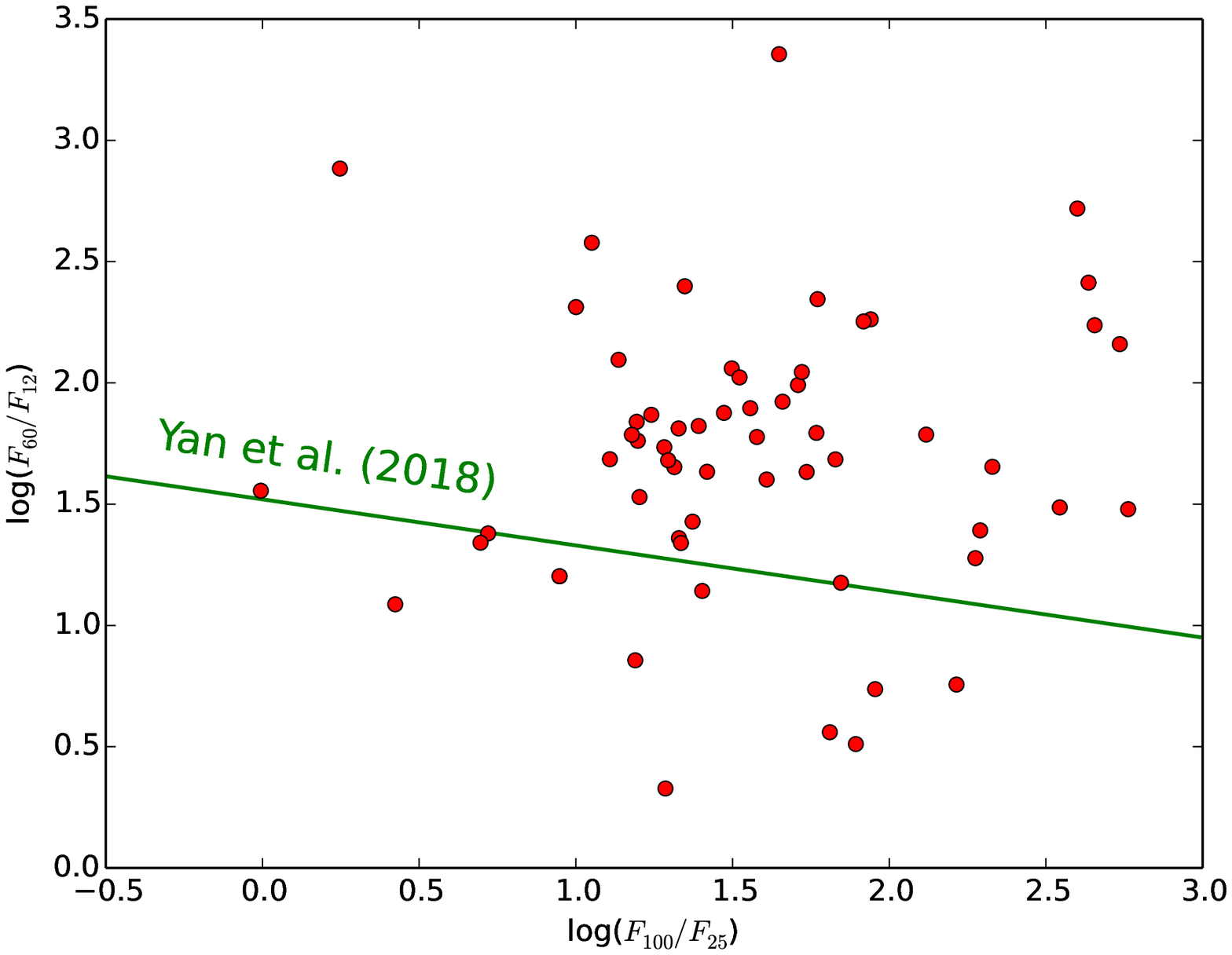}
\caption{The IRAS color-color diagrams for 59 sources. The solid green lines in the left-hand panel show the criteria of WC89 \citep{1989ApJ...340..265W}, and the UCH{\sc ii} regions lie in the upper right corner.
In the right-hand panel, the green line shows an improved H{\sc ii} region selection criterion \citep{2018MNRAS.476.3981Y}, and the H~{\sc ii} regions lie above the green line.\label{fig:iras}\\}
\end{figure}

The majority of the IRAS counterparts (93\%; 55/59) has a bolometric luminosity greater than 10$^3L_\odot$, indicating that they are probably associated with high-mass star forming regions.
About 83\% (49/59) of the IRAS counterparts have $F_{12}<F_{25}<F_{60}<F_{100}$, which is characteristic of YSOs. 
Figure \ref{fig:iras} shows IRAS color-color diagrams in which the youngest sources have the steepest spectra and are located toward the top right corner.  As a source evolves, the infrared colors become bluer and sources shift toward the lower left corner. The green lines in the left panel depict the criteria of \citet[][hereafter WC89]{1989ApJ...340..265W} which identify sources likely to be asscociated with an ultra-compact H~{\sc ii} (UCH~{\sc ii}) region, ($log(F_{25}/F_{12})\geqslant0.57$ and $log(F_{60}/F_{12})\geqslant1.3$). 
The left-hand panel in Fig.~\ref{fig:iras} shows that a significant fraction (61\% ; 36/59) of sources meet the WC89 criteria region.  This is a lower fraction than is observed for 6.7 GHz  class II methanol masers \citep[87\%;][]{2003ChJAA...3...49X}. 
This seems to be consistent with the maser-based evolutionary sequence for high-mass star formation, which suggests that while both class I methanol masers and 6.7 GHz class II methanol masers can occur during the UCH~{\sc ii} phase, the class I methanol masers are thought to disappear earlier than those of class II and so have less overlap with the UCH~{\sc ii} phase \citep{2010MNRAS.401.2219B}.
However, there are a number of potential biases that may impact our analysis. First, the WC89 criteria are for UCH~{\sc ii} candidates, but many of them have been found to be HMPOs rather than UCH{\sc ii} regions \citep[e.g.,][]{1996AA...308..573M,2002ApJ...566..931S}. Secondly, the beam sizes of 6.7 GHz maser observations (e.g., 3$\arcmin$--6$\arcmin$) were usually much larger than those (0.$\arcmin$5--1$\arcmin$) of 44 or 95 GHz maser observations, and hence there were more likely to be two or more massive YSOs in different evolutionary stages within the beam of 6.7 GHz maser observations. 
The green line in the right panel of Fig.~\ref{fig:iras} represents an improved H~{\sc ii} region selection criterion \citep{2018MNRAS.476.3981Y}, and we find that 48 sources satisfy this. 
Note that in applying the \cite{2018MNRAS.476.3981Y} criterion, we did not exclude sources with $Q_{60}=1$ or $Q_{100}=1$ as is suggested because under those circumstances the number of sources remaining is too small for a statistical analysis.

\begin{figure}
\center
\includegraphics[scale=0.4]{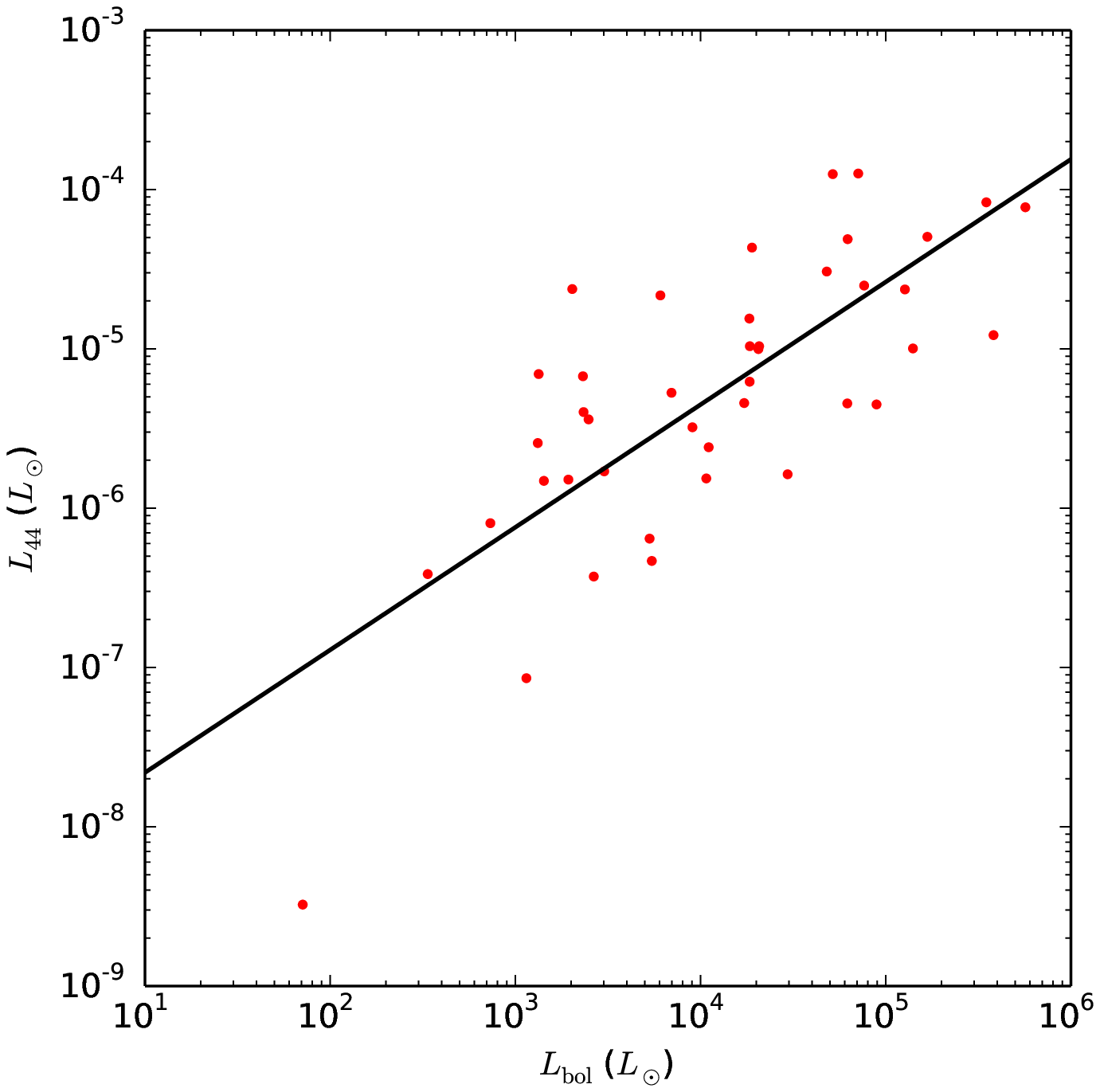}
\includegraphics[scale=0.4]{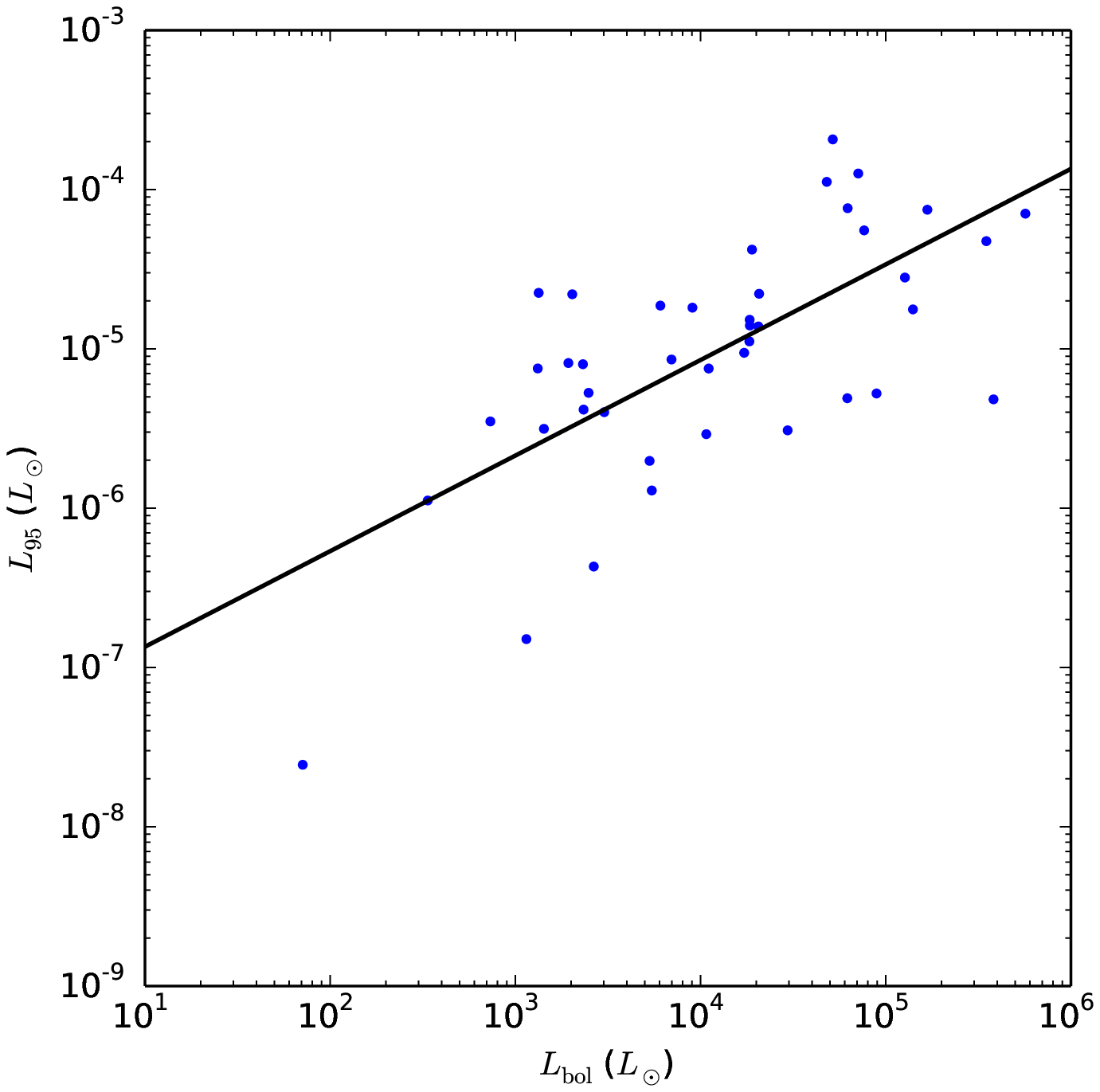}
\caption{The luminosity of class I methanol masers (left: 44 GHz; right: 95 GHz) versus the luminosity of IRAS counterparts for 42 sources. The black line depicts the best linear least-squares fit. \label{fig:LL}\\}
\end{figure}

After the methanol masers with an unreliable flux density are removed, a total of 42 sources that have both 44 and 95 GHz class I methanol maser emission and an IRAS source counterpart.
Fig.\ref{fig:LL} shows a plot of the isotropic maser luminosity as a function of the bolometric luminosity for the 44 GHz (left) and 95 GHz (right) methanol masers in our sample. 
The isotropic luminosities of the 44 and 95 GHz methanol masers ($L_{44}$ and $L_{95}$) can be calculated from the integrated flux density $\int S_v dv$ and the distance to the source \citep[see, e.g.,][]{2011ApJS..196...21B}. 
The black line in each panel marks the best linear fit to the data.
The best fit for the 44 GHz methanol masers is $log (L_{\rm 44}) =(0.77\pm0.11)\times log (L_{\rm bol})-(8.43\pm0.45)$ with a Pearson coefficient of 0.75. The best fit for the 95 GHz methanol masers is $log (L_{\rm 95}) =(0.60\pm0.10) \times log (L_{\rm bol})-(7.49\pm0.43)$ with a Pearson coefficient of 0.68.
Both $L_{44}$ and $L_{95}$ have a similar correlation with $L_{\rm bol}$, but the $L_{44}$ shows a slightly better correlation with $L_{\rm bol}$ than $L_{95}$.

\begin{figure}
\center
\includegraphics[scale=0.8]{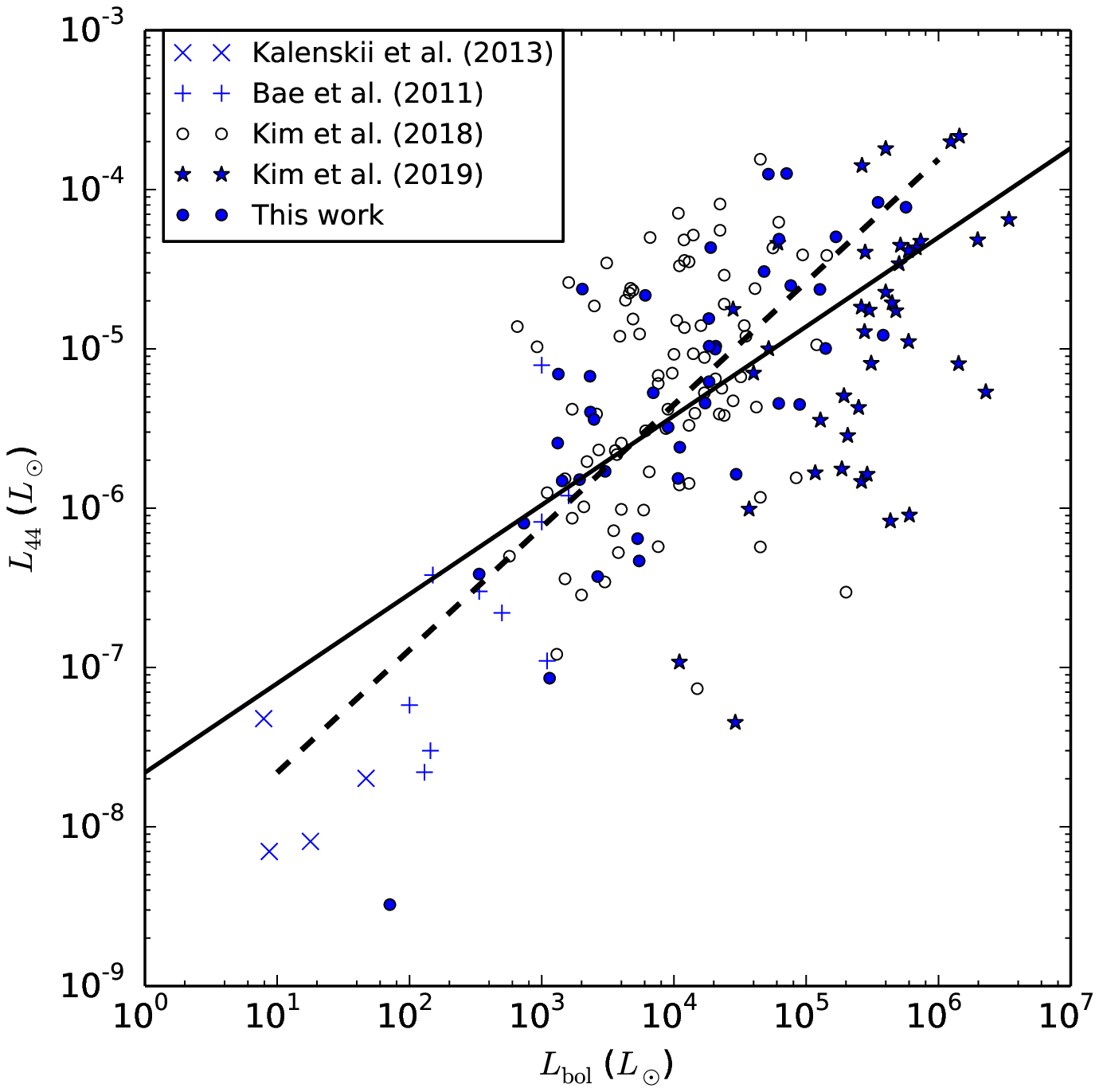}
\caption{ Isotropic maser luminosity as a function of the bolometric luminosity of the central star for 44 GHz methanol masers. Crosses, pluses, open circles, pentagrams, and solid circles represent low-mass YSOs \citep{2013ARep...57..120K}, intermediate-mass YSOs \citep{2011ApJS..196...21B}, high-mass protostellar object candidates from RMS sources \citep{2018ApJS..236...31K}, UCH{\sc ii} regions \citep{2019ApJS..244....2K} and our sample. The dashed and solid lines are the best linear least-squares fit for our sample and for all objects. 
\label{fig:LL2}\\}
\end{figure}

For 44 GHz methanol masers, the relation between $L_{44}$ and $L_{\rm bol}$ has been studied toward low-mass YSOs \citep[$L_{\rm bol}<10^2 L_{\odot}$;][]{2013ARep...57..120K}, intermediate-mass YSOs \citep[$L_{\rm bol}\sim 10^2-10^3 L_{\odot}$;][]{2011ApJS..196...21B}, HMPO candidates ($L_{\rm bol}>10^3 L_{\odot}$) from Red Midcourse Space Experiment Sources \citep{2018ApJS..236...31K}, and UCH~{\sc ii} regions \citep{2019ApJS..244....2K}.
Combining the data from these previous studies with our sample, the best fit for all sources is $log (L_{\rm 44}) =(0.56\pm0.05)\times log (L_{\rm bol})-(7.66\pm0.22)$ with a Pearson coefficient of 0.63. Fig. \ref{fig:LL2} shows the isotropic maser luminosity as a function of the bolometric luminosity for all objects. 
The bolometric luminosity is a good measure of the central object mass. The correlation between maser luminosity and bolometric luminosity may arise when sources with higher mass sources drive more powerful outflows, or at least outflows that result in a larger volume of gas that produces class I methanol masers, where those outflows interact with the surrounding molecular environment \citep[e.g.,][]{2011ApJS..196...21B,2019ApJS..244....2K}.

\subsection{44 GHz methanol masers occur at different evolutionary stages}

It is widely accepted that the earliest stages of high-mass star formation often take place while the source is embedded within an IRDC \citep{2006ApJ...641..389R}. Investigation of molecular gas within IRDCs compared to HMPOs, shows that as the sources evolve, the temperatures increase and the densities and masses rise. The next evolutionary phase is associated with UCH~{\sc ii} regions, which are produced by young massive stars as they reach the main sequence \citep[e.g.,][]{2018ARAA..56...41M}. Thus molecular clumps associated with IRDCs, HMPOs, and UCH~{\sc ii}s represent an approximate evolutionary sequence of massive star formation.

\cite{2006ApJ...638..241E} found that the infrared color of sources that are only associated with class I methanol masers are redder than the color of sources associated with both class I and class II methanol masers, indicating that the class I methanol masers trace an earlier stage than class II methanol masers. 
On the other hand, a few previous studies suggested that the detection rate of 44 GHz class I methanol masers increases as the central objects evolve \citep[e.g.,][]{2011ApJS..196...21B,2019ApJS..244....2K}.
These studies indicate that 44 GHz class I methanol masers can be associated with both early and late evolutionary stages of high-mass star formation, but the masers are more frequently detected in later evolutionary stages.
Theoretical prediction and interferometric observations \citep[e.g.,][]{2014MNRAS.439.2584V} also found that class I methanol maser could be pumped by expanding H~{\sc ii} regions, demonstrating that some class I methanol masers are associated with later stages of high-mass star formation.  
We would like to determine whether the current sample contains class I masers likely associated with both younger and older SFRs.

We investigated the infrared emission detected by the \textit{all-sky Wide-field Infrared Survey Explorer} \citep[\textit{WISE}\footnote{\url{http://irsa.ipac.caltech.edu/Missions/wise.html}};][]{2010AJ....140.1868W} toward each of our target sources.  The \textit{WISE} images can be used to determine which class~I methanol maser sources may be associated with an infrared dark background or mid-infrared morphologies suggesting H{\sc ii} regions. 

WISE used a 40 cm telescope to image the entire sky in four mid-IR bands at 3.4, 4.6, 12 and 22 $\mu$m. From 3.4 to 22 $\mu$m, WISE achieved sensitivities for point sources of 0.08, 0.11, 1, and 6 mJy and angular resolution of 6$\arcsec$.1, 6$\arcsec$.4, 6$\arcsec$.5 and 12$\arcsec$.0 in the four bands, respectively. Mid-IR emission can be used to trace star-forming activity through polycyclic aromatic hydro-carbon emission, seen in the WISE 12$\mu$m band, and through the correlation with thermal emission from warm dust, predominantly see in the WISE 22 $\mu$m band.

\begin{figure}
\plotone{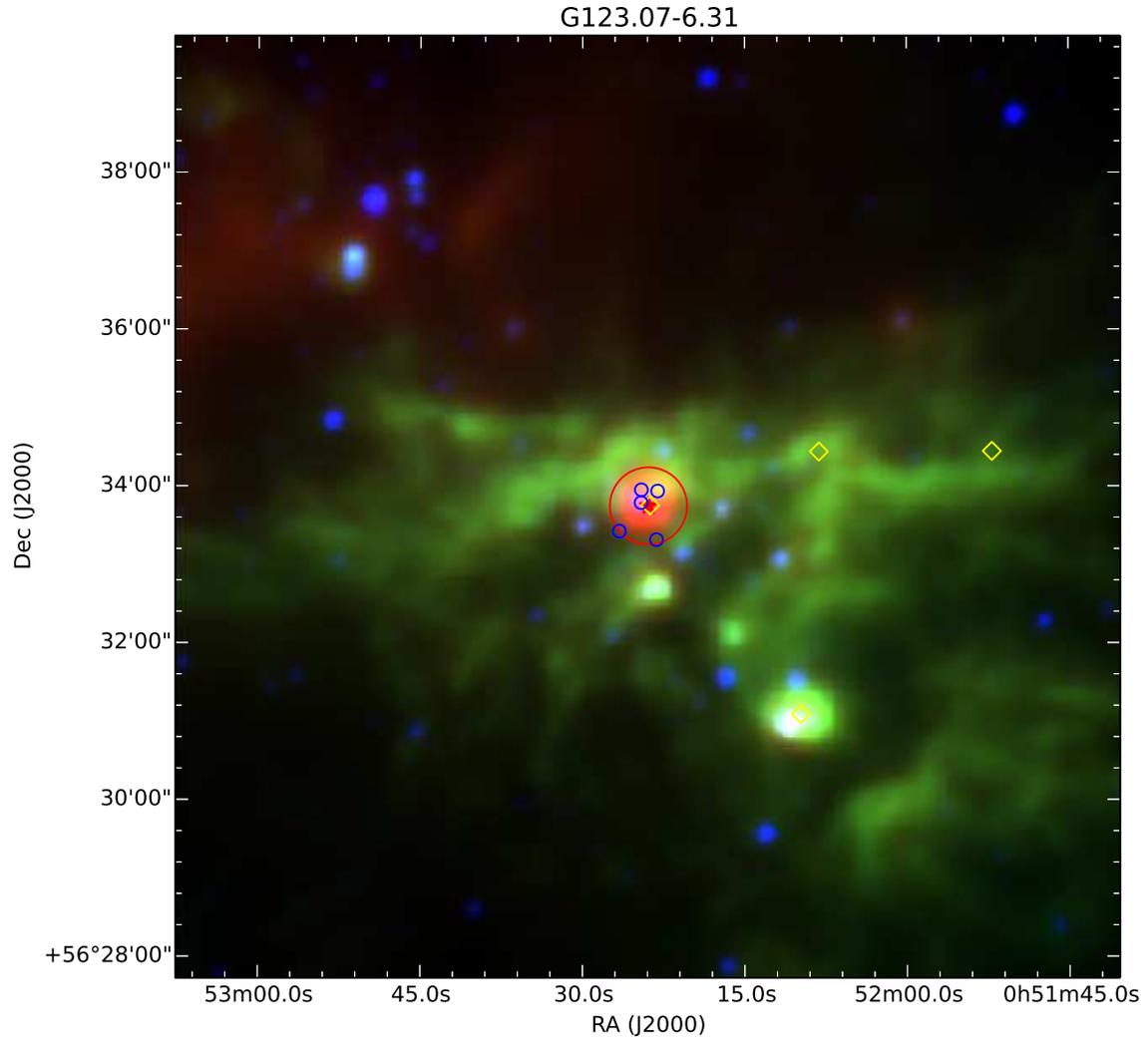}
\caption{Red, green and blue colors in each figure represent the WISE 22, 12, and 4.6 $\mu$m emission, respectively. The red star indicates the pointing center of the observations, the large red circle indicates the approximate beam size $\sim$1$\arcmin$ at 44 GHz (the beam size at 95 GHz is about a half of that at 44 GHz), the blue dots show the location of WISE point sources within the beam, the yellow diamonds show the location of IRAS sources in this 12$\arcmin$ $\times$ 12$\arcmin$ region.\\
(The complete figure set (144 images) is available in the online material.)\label{fig:infra}}
\end{figure}

We inspected the 12$\arcmin$ $\times$ 12$\arcmin$ three-color images for each of our target maser sources (see Figure \ref{fig:infra} as an example).  
From our sample, we found that some class I methanol masers could be associated with IRDCs and some could be associated with H~{\sc ii} regions, suggesting that there are both younger and older class I methanol maser sources during star formation.

\subsubsection{Infrared dark clouds}  \label{sec:sec_irdc}

The IRDCs are seen as a dark silhouette against the bright background emission in images at wavelengths between 7 and 25 $\mu$m  \citep[e.g.,][]{2006ApJ...639..227S}.  Investigations of molecular line and dust continuum emission shows that they are regions of dense, cold gas and have high column densities, consistent with the expectations for the initial conditions for high-mass star formation \citep[e.g.,][]{2006AA...450..569P,2006ApJ...641..389R}.

\cite{2009AA...505..405P} compiled an unbiased sample of candidate IRDCs (11303 in total) in the 10$^\circ < \mid l \mid <65 ^\circ$, $\mid b \mid < 1^\circ$ region of the Galactic plane using Spitzer 8 $\mu$m extinction. Subsequently, 76($\pm$19)\% of these cataloged SDCs were confirmed through association with a peak in Herschel column density maps constructed from 160 $\mu$m and 250 $\mu$m data from Herschel Galactic plane survey Hi-GAL \citep{2016AA...590A..72P}.

In the WISE three-color images, 98\% (125/128) have at least one WISE point source within 30$\arcsec$ (about half of the beam size at 44 GHz) of the pointing center.  Two of the exceptions are G208.97$-$19.37 (Orion-KL) and BGPS2152 (M17), where the three-color images are saturated and only one source (BGPS2054) has no close WISE point source in the AllWISE catalog.
For the majority of sources with a 95 GHz class~I methanol maser detection, we can surmise that the emission is located within the 15$\arcsec$ of the targeted center (because the sensitivity of the observations drops very rapidly at larger separations).  When Orion-KL and M17 are excluded, 25\% (31/126) of the targeted locations have no WISE point source within 15$\arcsec$ (about half of the beam size in 95 GHz) of the pointing center.  We examined the WISE three-color images of these 31 masers and found that some of them show an absence of bright 22 $\mu$m emission, perhaps indicating the presence of an IRDC.

With the exception of IRDC18223-3 (where the observed target is a known IRDC), for the 31 sources without nearby ($\sim$15$\arcsec$) WISE counterparts, we cross-matched with the Herschel-confirmed SDC catalog, finding that 5 of them may be associated with IRDCs (See Table \ref{tab:irdc}).
The ratios, $S_{\rm pk,95}$/$S_{\rm pk,44}$, for the masers that may be associated with IRDCs are no greater than one, but for IRDC18223-3, the peak emission of the 95 GHz methanol maser is stronger than that of the 44 GHz counterpart of $S_{\rm pk,95}$/$S_{\rm pk,44}$ $\sim$ 1.2.

\subsubsection{masers in the vicinity of H {\sc ii} regions} \label{sec_hii}

\cite{2014ApJS..212....1A} compiled a catalog of more than 8000 Galactic H{\sc ii} regions and candidates by visually and automatically searching for their characteristic mid-IR morphology using WISE data.
Approximately 1500 of the mid-infrared selected sample have been detected in RRL emission and are thus confirmed to be H~{\sc ii} regions. 
Furthermore, as a part of the Green Bank Telescope H~{\sc ii} Region Discovery Survey, new RRL observations are being undertaken \citep{2015ApJ...810...42A,2018ApJS..234...33A}, expanding the catalog of confirmed H~{\sc ii} regions.

We cross-matched the class I methanol masers targeted in the current observations with the WISE catalog of Galactic H~{\sc ii} regions v2.0 from \url{http://astro.phys.wvu.edu/wise/}\citep[see also][]{2018ApJS..234...33A}.
The WISE name, coordinates, the approximate circular radius encompassing the WISE mid-IR emission and corresponding RRL velocity information were extracted from the catalog.
We consider the class I methanol maser to be possibly associated with a H~{\sc ii} region when it meets the following criteria:
(1) The angular separation between maser and H~{\sc ii} region is smaller than the radius of the corresponding H~{\sc ii} region. 
(2) To avoid including the effects from other complex environments, the angular separation between maser and H~{\sc ii} region is no greater than 60$\arcsec$ (approximately the beam size at 44 GHz).  It should be noted that this criterion may result in not identifying cases where the maser is located at the edge of an evolved and extended H~{\sc ii} region. 
(3) The H~{\sc ii} region has detected RRL emission, and the absolute difference between the velocity of the RRL and class I methanol maser is less than $\sim$ 10 \kms \citep{2014ApJS..212....1A}.

The criteria identified a total of 22 class I methanol masers (see Table \ref{tab:HII}) that may be associated with H~{\sc ii} regions, all of which also have BGPS 1.1 mm emission.  
There are multiple WISE H~{\sc ii} region counterparts within 30$\arcsec$ (about half of the beam size in 44 GHz) of two masers (BGPS1116 and BGPS3155), suggesting that they are part of a larger H~{\sc ii} region complex.
We listed all the possible WISE H~{\sc ii} regions, because when the accurate position of maser is unknown, we only interested in which masers may be associated with H~{\sc ii} regions rather than which specific H~{\sc ii} region it is associated with.  
Table \ref{tab:HII} lists the names of the methanol masers, the possible associated WISE H~{\sc ii} regions, the angular separation between the BGPS source and H{\sc ii} region, the angular sizes of the H~{\sc ii} regions, the velocity range of class I methanol masers and the velocity of the RRLs from H~{\sc ii} regions, as well as the ratio of the peak flux density for the two class~I methanol transitions. 
For the source with an unreliable flux density or a 95 GHz nondetection, the ratio of the peak flux density between the two transitions is not listed. 
The ratios, $S_{\rm pk,95}$/$S_{\rm pk,44}$, for the masers that may be associated with H~{\sc ii} regions are all lower than one.

It should be noted that a large number of observations have shown that CO outflows are common toward both HMPOs and UCH~{\sc ii}s \citep[e.g.,][]{1996ApJ...472..225S,2001ApJ...552L.167Z,2002ApJ...566..931S}, and numerous YSOs are observed at the peripheries of compact or extended H~{\sc ii} regions \citep[e.g.,][]{2008A&A...482..585D,2009A&A...494..987P}.
Thus it is difficult to differentiate whether class I methanol masers are produced by the expansion of H{\sc ii} regions or the outflows of associated YSOs, based on low-resolution radio and IR data. 
High-resolution observations are necessary to clarify the issue.

Four sources (BGPS1954, BGPS3026, BGPS3307, and BGPS4933) which may be associated with a H~{\sc ii} region and also a nearby IRDC (within $\sim$ 1 $\arcmin$). 
The VLA observation of BGPS3307 (see Sec.\ref{sec:44_compare}) shows that the maser components are located at the interface between a H~{\sc ii} region and an IRDC.
For the other three sources, the exact position of the maser emission is unknown, and hence we cannot determine whether they are associated with an IRDC.

\section{Summary} \label{sec:sum}

We have used the three KVN antennas in single-dish mode to simultaneously observed the 44 and 95 GHz class I methanol masers toward 144 sources with previous 95 GHz detections. The main results are listed below.\\
1. 128 44 GHz and 111 95 GHz methanol were detected, corresponding to detection rates of 89\% and 77\%, respectively. 
This is the first reported detection of 44~GHz class~I methanol masers for 106 sites. \\
2. Through comparison with previous observations with PMO 13.7m new observations, no clear evidence was found for variability in 95 GHz methanol masers.\\
3. The 44 and 95 GHz methanol masers show a strong correlation in peak velocity, peak flux density, and integrated flux density. The peak flux density ratio $S_{\rm pk,95}$/$S_{\rm pk,44}$ ranges from 0.1 to 2.8, and the best fit for $S_{\rm pk,95}$/$S_{\rm pk,44}$ in our sample is $S_{\rm pk,95}$=(0.49$\pm$0.01)$S_{\rm pk,44}$+(1.88$\pm$0.66).
We found that the peak flux density ratio $S_{\rm pk,95}$/$S_{\rm pk,44}$ decreases as the 44 GHz peak flux density increases.
No 95 GHz methanol maser are stronger than the 44 GHz counterpart when the peak flux density of the 44 GHz maser is stronger than 40 Jy, and in only $\sim$ 15\% of the sources in our sample dose the 95 GHz peak flux density surpass that of the 44 GHz counterpart.\\
4. Class I methanol masers occur at early stage and more evolved stage of high-mass star formation.
There are some class~I methanol masers in our sample that could be associated with IRDCs and others with  H{\sc ii} regions.\\

\acknowledgments
We acknowledge the anonymous referee for helpful comments that have improved this paper.
We are grateful to the staff of the KVN observatory and PMO observatory for their assistance during the observations.
We also thank Dr. Yan Gong and Dr. Qingzeng Yan for helpful discussions and suggestions.
This work was sponsored by the MOST under grand No. 2017YFA0402701, the National Natural Science Foundation of
China (grant numbers: 11933011, 11873019 and 11673066), and the Key Laboratory for Radio Astronomy, CAS.
A.M.S was supported by the Ministry of Science and Education, FEUZ-2020-0030.
W.Y. thanks her parents for feeding her tons of foods during the COVID-19 outbreak period.

%
\vspace{5mm}
\facilities{KVN, PMO, WISE, IRAS}

\software{GILDAS/CLASS \citep{2005sf2a.conf..721P,2013ascl.soft05010G}, Numpy \citep{2011CSE....13b..22V}, Matplotlib \citep{2007CSE.....9...90H} and APLpy \citep{2012ascl.soft08017R}.}

\clearpage
\begin{deluxetable}{llrccccc}
\tablecaption{Target sources.\label{tab:144_obs}}
\tablecolumns{8}
\tablenum{1}
\tablewidth{0pt}
\tablehead{
\colhead{Source} & \colhead{R.A.} & \colhead{decl.}  & \multicolumn{2}{c}{Detection} & \colhead{New 44 GHz}& \colhead{$\sigma_{44}$} & \colhead{$\sigma_{95}$}\\
\cline{4-5}
\colhead{} & \colhead{(J2000)} & \colhead{(J2000)}  & \colhead{44 GHz}  & \colhead{95 GHz} & \colhead{detection?} & \colhead{(Jy)}  & \colhead{(Jy)} } 
\startdata
G123.07$-$6.31$^\ddagger$ & 00 52 23.9 & 56 33 45 & Y & Y & Y & 0.6  & 0.6\\
BGPS7351 & 02 25 30.8 & 62 06 18 & Y & Y & N & 0.4  & 0.4\\
G208.97$-$19.37 & 05 35 15.5 & $-$05 20 41 & Y & Y & Y & 0.6  & 0.7\\
G208.77$-$19.24 & 05 35 21.8 & $-$05 07 37 & Y & Y & Y & 0.3  & 0.5\\
G173.58$+$2.44 & 05 39 27.5 & 35 40 43 & Y & Y & Y & 0.3  & 0.5\\
\enddata
\tablecomments{The sources marked with a dagger have unreliable flux densities due to poor weather. The sources with previous interferometric observations are marked with an asterisk. Ten sources were assigned as maser candidates \citep{2017ApJS..231...20Y} and are noted with a double dagger. Sources marked with a dagger and an asterisk are not given in this portion of table, they can be found in the full online table.\\
Reference for cross-matching new 44 GHz detections: \cite{2019AJ....158..233L}.\\
References for interferometric observations: 1. \cite{2004ApJS..155..149K} 2. \cite{2016ApJS..222...18G} 3. \cite{2017ApJS..233....4R}\\
(This table is available in the online material. A portion is shown here for guidance.)}
\end{deluxetable}

\begin{deluxetable}{lccccc}
\tablecaption{Telescope Parameters\tablenotemark{a}. \label{tab:tele_para}}
\tablecolumns{6}
\tablenum{2}
\tablewidth{0pt}
\tablehead{
\colhead{Telescope} &
\colhead{Frequency\tablenotemark{b}} &
\colhead{Beam\tablenotemark{c}} & \colhead{$\eta_{\rm mb}$\tablenotemark{d}} & \colhead{$\eta_{\rm A}$\tablenotemark{d}} & \colhead{$f$\tablenotemark{e}}  \\
\colhead{} & \colhead{(GHz)} &
\colhead{(arcsec)} & \colhead{} & \colhead{} & \colhead{(Jy K$^{-1}$)}
}
\startdata
KVN Yonsei & 43 & 63 & 0.48 & 0.64 & 15.5 \\
           & 86 & 32 & 0.41 & 0.53 & 18.7 \\
KVN Ulsan  & 43 & 63 & 0.47 & 0.62 & 16.0 \\
           & 86 & 32 & 0.44 & 0.57 & 17.4 \\
KVN Tamna  & 43 & 63 & 0.49 & 0.64 & 15.5 \\
           & 86 & 32 & 0.44 & 0.57 & 17.4 \\
\enddata
\tablenotetext{a}{Telescope parameters can be found in the KVN status report: \\
\url{https://radio.kasi.re.kr/kvn/status_report_2016/aperture_efficiency.html}}
\tablenotetext{b}{Frequency where the beam size and efficiencies were measured.}
\tablenotetext{c}{Full width at half-maximum (FWHM) of the main beam.}
\tablenotetext{d}{Main-beam efficiency $\eta_{\rm mb}$ and aperture efficiency $\eta_{\rm A}$. In each receiver band, $\eta_{\rm A}$ is nearly the same, but $\eta_{\rm mb}$ is inversely proportional to the square of frequency. }
\tablenotetext{e}{Scaling factor for converting the KVN raw data into spectra in the flux density scale for the maser lines, which includes the quantization correction factor (1.25).}
\end{deluxetable}

\begin{deluxetable}{lcccccccc}
\tablecaption{Comparison of 95 GHz PMO Reobservations. \label{tab:f_comparison}}
\tablecolumns{9}
\tablenum{3}
\tablewidth{0pt}
\tablehead{
\colhead{Source} & \multicolumn{2}{c}{New PMO obs.}  & \multicolumn{3}{c}{Previous PMO obs.} & \multicolumn{2}{c}{KVN obs.} &\colhead{Notes}\\
\cmidrule(r){2-3} \cmidrule(r){4-6} \cmidrule(r){7-8}
\colhead{} & \colhead{Peak (Jy)} &  \colhead{rms (Jy)}&\colhead{Date} & \colhead{Peak (Jy)} &  \colhead{rms (Jy)} & \colhead{44 GHz} & \colhead{95 GHz} & \colhead{}}
\startdata
BGPS1917 &9 &0.9 &201605,201405 &9 & 0.7 &23 &18 &a\\
BGPS2011 &8 &0.8 &201406        &8 & 1.0 &13 &N  &b\\
BGPS2720 &5 &0.9 &201406        &7 & 0.9 &12 &N &b\\
BGPS2784 &N &1.0 &201605,201306,201206 &5 & 0.8 &4 &N &c\\
BGPS3319 &N &0.9 &201206        &6 & 1.3 &N  &N &d\\
BGPS3322 &5 &0.9 &201206        &8 & 1.5 &11 &N &b\\
BGPS4063 &N &1.0 &201406,201307 &5 & 1.1 &2  &N &c\\
G031.013+00.078 &N &1.0 &201103 &4 & 0.8 &6  &N &c\\
BGPS4252 &60 &1.3 &201206       &75 & 1.6 &280 &130 &a\\
BGPS7125 &5 &0.8 &201505,201205 &6 & 0.7 &26 &N &b\\
BGPS7252 &9 &0.8 &201205        &11 & 0.6 &9 &2 &b\\
BGPS7322 &4 &0.8 &201205        &5 & 0.9 &6 &N &b\\
BGPS7351 &4 &0.8 &201605,201205 &5 & 0.8 &2 &2 &b\\
\enddata
\tablecomments{\\
Column 1: the source name. Columns 2--3: the peak flux density and 1$\sigma_{\rm rms}$ noise of the new PMO observations. Columns 4--6: the observational dates, the peak flux density, and 1$\sigma_{\rm rms}$ noise of the previous PMO observations. Columns 7--8: the peak flux density of 44 and 95 GHz methanol masers of KVN observations. N indicates a nondetection. Note that the velocity resolutions of KVN data at both 44 and 95 GHz are $\sim$ 0.2 \kms .
Column 9: the comments of each source, the detailed situations are listed below:\\
a. The peak flux density is affected by a different velocity resolution.\\
b. The maser could be out of the beam at 95 GHz or may be at the edge of the beam.\\
c. The source could have week 95 GHz emission, but cannot be distinguished from noise.\\
d. The source could have no 95 GHz emission, the previous detection might be a false detection.\\}
\end{deluxetable}

\begin{deluxetable}{lcrrcrc}
\tablecaption{Parameters of Gaussian fits to 44 GHz methanol maser detections.\label{tab:44_para}}
\tablecolumns{7}
\tablenum{4}
\tablewidth{0pt}
\tablehead{
\colhead{Source} & \colhead{Components} & \colhead{$V_{\rm pk,44}$} & \colhead{$\Delta V_{44}$} & \colhead{$S_{\rm pk,44}$} & \colhead{$S_{44}$} &  \colhead{${\int S_{44}dv}$} \\
\colhead{} & \colhead{} & \colhead{(\kms)}  & \colhead{(\kms)}  & \colhead{(Jy)}  & \colhead{(Jy \kms)}  & \colhead{(Jy \kms)} } 
\startdata
G123.07$-$6.31     & a         & $-$33.20  (0.21)  & 2.15  (0.21)  & 0.8 (0.6)     & 1.9    (0.2)  & 8.8    \\
G123.07$-$6.31     & b         & $-$30.34  (0.21)  & 2.88  (0.21)  & 2.2 (0.6)    & 6.9    (0.2)  & \nodata\\
BGPS7351         & a         & $-$45.41  (0.24)  & 1.40  (0.49)  & 0.8 (0.4)    & 0.9    (0.3)  & 5.7 \\
BGPS7351         & b         & $-$42.25  (0.14)  & 2.92  (0.37)  & 1.9 (0.4)    & 4.8    (0.5)  & \nodata\\
G208.97$-$19.37    & a         & 7.26    (0.05)  & 0.30  (0.24)  & 1.8 (0.6)     & 0.6    (0.3)  & 6.8 \\
G208.97$-$19.37    & b         & 8.62    (0.11)  & 1.10  (0.38)  & 3.1 (0.6)   & 3.7    (1.0)  & \nodata\\
G208.97$-$19.37    & c         & 10.00   (0.13)  & 0.85  (0.42)  & 2.8 (0.6)    & 2.5    (1.0)  & \nodata\\
G208.77$-$19.24    & \nodata   & 11.22   (0.05)  & 0.27  (0.07)  & 1.5 (0.3)    & 0.4    (0.1)  & 0.4 \\
G173.58+2.44     & \nodata   & $-$17.03  (0.02)  & 0.63  (0.06)  & 4.1 (0.3)    & 2.8    (0.2)  & 2.8 \\
\enddata
\tablecomments{Column 1: Source name. Column 2: The component(s) of each source at 44 GHz.
Columns 3--6: The velocity at peak $V_{\rm pk,44}$, the line width $\Delta V_{44}$, the peak flux density $S_{44}$, the integrated flux density $S_{44}$ and corresponding fitting error for each of the maser features that have been estimated from Gaussian fits. Column 7: The total integrated flux density ${\int S_{44}dv}$ of the maser spectrum obtained by adding the integrated flux density of all maser features in the source.\\
(This table is available in the online material. A portion is shown here for guidance.)}
\end{deluxetable}

\begin{deluxetable}{lcrrcrc}
\tablecaption{Parameters of Gaussian fits to 95 GHz methanol maser detections.\label{tab:95_para}}
\tablecolumns{10}
\tablenum{5}
\tablewidth{0pt}
\tablehead{
\colhead{Source} & \colhead{Components} & \colhead{$V_{\rm pk,95}$} & \colhead{$\Delta V_{95}$} & \colhead{$S_{\rm pk,95}$} & \colhead{$S_{95}$} &  \colhead{${\int S_{95}dv}$} \\
\colhead{} & \colhead{} & \colhead{(\kms)}  & \colhead{(\kms)}  & \colhead{(Jy)}  & \colhead{(Jy \kms)}  & \colhead{(Jy \kms)} } 
\startdata
G123.07$-$6.31      & a        &  $-$32.94  (0.20)    & 1.33  (0.20)    & 2.8 (0.6)    & 3.9  (0.4)   & 23.0   \\
G123.07$-$6.31      & b        &  $-$31.58  (0.20)    & 0.92  (0.20)    & 4.8 (0.6)    & 4.7  (0.4)   & \nodata\\
G123.07$-$6.31      & c        &  $-$30.52  (0.20)    & 1.12  (0.20)    & 6.1 (0.6)    & 7.3  (0.4)   & \nodata\\
G123.07$-$6.31      & d        &  $-$29.30  (0.20)    & 1.20  (0.20)    & 3.9 (0.6)    & 5.0  (0.4)   & \nodata\\
G123.07$-$6.31      & e        &  $-$27.89  (0.20)    & 1.32  (0.20)    & 1.5 (0.6)    & 2.1  (0.4)   & \nodata\\
BGPS7351          & a        &  $-$45.56  (0.16)    & 0.73  (0.25)    & 0.8 (0.4)   & 0.7  (0.2)   & 4.2 \\
BGPS7351          & b        &  $-$43.79  (0.09)    & 1.21  (0.24)    & 1.6 (0.4)   & 2.1  (0.3)   & \nodata\\
BGPS7351          & c        &  $-$42.49  (0.08)    & 0.54  (0.15)    & 1.2 (0.4)  & 0.7  (0.2)   & \nodata\\
BGPS7351          & d        &  $-$41.60  (0.04)    & 0.41  (0.12)    & 1.7 (0.4)   & 0.7  (0.2)   & \nodata\\
G208.97$-$19.37     & a        &  8.09    (0.20)    & 0.95  (0.36)    & 1.4 (0.7)   & 1.4  (0.4)   & 7.5 \\
G208.97$-$19.37     & b        &  8.93    (0.07)    & 0.33  (0.24)    & 1.7 (0.7)    & 0.6  (0.4)   & \nodata\\
G208.97$-$19.37     & c        &  9.95    (0.05)    & 1.20  (0.16)    & 4.3 (0.7)    & 5.5  (0.6)   & \nodata\\
G208.77$-$19.24     & a        &  11.25   (0.06)    & 0.55  (0.31)    & 1.8 (0.5)  & 1.0  (0.4)   & 1.4 \\
G208.77$-$19.24     & b        &  11.95   (0.07)    & 0.30  (0.31)    & 1.3 (0.5)   & 0.4  (0.3)   & \nodata \\
G173.58+2.44      & \nodata  &  $-$17.23  (0.08)    & 0.82  (0.19)    & 1.8 (0.3)  & 1.5  (0.3)   & 1.5 \\
\enddata
\tablecomments{Column 1: Source name. Column 2: The component(s) of each source at 95 GHz.
Columns 3--6: The velocity at peak $V_{\rm pk,95}$, the line width $\Delta V_{95}$, the peak flux density {$S_{\rm pk,95}$}, the integrated flux density $S_{95}$ and corresponding fitting error for each of the maser features that have been estimated from Gaussian fits. Column 7: The total integrated flux density ${\int S_{95}dv}$ of the maser spectrum obtained by adding the integrated flux density of all maser features in the source.\\
(This table is available in the online material. A portion is shown here for guidance.)}
\end{deluxetable}

\begin{deluxetable}{lcccccccccccc}
\tablecaption{The IRAS counterparts for 44-GHz class I methanol masers. \label{tab:iras}}
\tablecolumns{13}
\tablenum{6}
\tablewidth{0pt}
\tablehead{
\colhead{Source} & \colhead{Angular Sep.} &  \colhead{IRAS}&\colhead{$F_{12}$} & \colhead{$F_{25}$} &\colhead{$F_{60}$} & \colhead{$F_{100}$} &\colhead{$Q_{12}$} & \colhead{$Q_{25}$} &\colhead{$Q_{60}$} & \colhead{$Q_{100}$} &\colhead{Distance} & \colhead{Luminosity}\\
\colhead{} & \colhead{(arcsec)} &  \colhead{}&\colhead{(Jy)} & \colhead{(Jy)} &\colhead{(Jy)} & \colhead{(Jy)} &\colhead{} & \colhead{} &\colhead{} & \colhead{} &\colhead{(kpc)} & \colhead{(10$^3L_\odot$)}}
\startdata
G123.07$-$6.31	&1	&00494+5617	&1.803	&13.36	&329.6	&1166.0	&1	&3	&3	&3	&2.82	&9.0\\
G208.77$-$19.24	&39	&05329$-$0508	&0.3883	&24.8	&297.1	&43.82	&1	&3	&1	&1	&0.42	&0.1\\
G173.58$+$2.44	 &1	&05361+3539	&1.182	&6.715	&29.15	&1310.0	&3	&3	&3	&1	&1.7	&2.7\\
G207.27$-$1.81	&0	&06319+0415	&78.44	&375.2	&958.8	&995.2	&3	&3	&3	&3	&1.3	&5.5\\
BGPS1062	 &33	 &17542$-$2447	&4.165	&4.401	&13.51	&344.3	&1	&1	&3	&1	&2.1	&1.2\\
BGPS1116	 &28	 &17545$-$2357	&11.48	&106.6	&793.9	&1667.0	&3	&3	&3	&3	&1.9	&8.2\\
BGPS1138	 &43	 &17571$-$2401	&24.43	&67.23	&12790.0	&26780.0	&1	&3	&1	&1	&2.3	&159.8\\
\enddata
\tablecomments{Column 1: the maser name. Column 2: the angular separation of KVN targeted center and possible associated IRAS source. Columns 3: IRAS name. Columns 4--7: the flux density in the four IRAS bands. Columns 8--11: the corresponding flux quality in the four \textit{IRAS} band. Column 12: distance for each of the target sources \citep{2017ApJS..231...20Y}, primarily using kinematic distances calculated from \cite{2014ApJ...783..130R}. Column 13: the calculated bolometric luminosity of \textit{IRAS} source \citep{2007AJ....133.1528C}.\\
(This table is available in the online material. A portion is shown here for guidance.)}
\end{deluxetable}

\begin{deluxetable}{ccc}
\tablecaption{Maser candidates that may be associated with an IRDC. \label{tab:irdc}}
\tablecolumns{3}
\tablenum{7}
\tablewidth{0pt}
\tablehead{
\colhead{Maser name} &\colhead{SDC name} &\colhead{$S_{\rm pk,95}$/$S_{\rm pk,44}$}}
\startdata
BGPS1584  & SDC011.081$-$0.532 &0.4 \\
BGPS2054  & SDC014.493$-$0.143 &\nodata \\
BGPS2718  & SDC020.731$-$0.055 &0.6 \\
BGPS3018  & SDC023.210$-$0.371 &1 \\
BGPS4557  & SDC030.811$-$0.110 &0.6 \\
\enddata
\end{deluxetable}

\begin{deluxetable}{cclcccc}
\tablecaption{Masers that may be associated with H~{\sc ii} regions. \label{tab:HII}}
\tablecolumns{7}
\tablenum{8}
\tablewidth{0pt}
\tablehead{
\colhead{Maser} &\colhead{Angular sep.} & \colhead{WISE H~{\sc ii}} & \colhead{Radius of H~{\sc ii}}& \colhead{Maser velocity range} & \colhead{$V_{\rm RRL}$} & \colhead{$S_{\rm pk,95}$/$S_{\rm pk,44}$}\\
 & \colhead{(arcsec)} & & \colhead{(arcsec)}  & \colhead{(\kms)}  & \colhead{(\kms)}  & }
\startdata
BGPS7501	    & 7    & G192.584$-$00.043    &	60	 & 6--12        & 7.5        & 0.6     \\
BGPS1116	    & 24   & G005.637$+$00.232      &	19	 & 6--10        & 6.3        & \nodata     \\
\nodata	    & 28   & G005.633$+$00.238      &	59	 & 6--10        & 6.3        & \nodata     \\
BGPS1954	    & 27   & G013.880$+$00.285      &  144	 & 49--53       & 51         & \nodata  \\  
BGPS2011    & 48   & G014.207$-$00.193    &  608  & 38--43       & 36.1       & \nodata  \\
BGPS2275	    & 18   & G016.360$-$00.211    &	33	 & 47--51       & 46.4       & \nodata     \\
BGPS2784	    & 2    & G021.386$-$00.255    &	57	 & 90--92       & 91.2       & \nodata     \\
BGPS3026	    & 13   & G023.264$+$00.077      &	60	 & 76--82       & 78.2       &  0.4    \\
BGPS3155	    & 19   & G023.708$+$00.174      &  171	 & 112--115     & 103.8      &  0.6   \\    
\nodata	    & 22   & G023.713$+$00.175      &	44	 & 112--115     & 103.8      &  0.6    \\
\nodata	    & 24   & G023.705$+$00.165      &	46	 & 112--115     & 103.8      &  0.6    \\
BGPS3183	    & 21   & G023.872$-$00.119    &	60	 & 71--79       & 73.8       &  0.5    \\
BGPS3337	    & 20   & G024.498$-$00.039    &	60	 & 107--112     & 108.1      &  0.7    \\
BGPS3474	    & 27   & G025.220$+$00.289      &	42	 & 45--47       & 42.4       &  0.9    \\
G024.920$+$00.085	& 21   & G024.923$+$00.079  &	42	 & 40--50       & 42.4       &  \nodata    \\
BGPS3307	    & 13   & G024.397$-$00.191    &	47	 & 58--62       & 54.7       &  0.8    \\
BGPS3774	    & 21   & G027.279$+$00.143      &	33	 & 30--34       & 36.3       &  0.7    \\
BGPS4014	    & 7    & G028.651$+$00.026      &	33	 & 100--105     & 102.4      &  0.6    \\
BGPS4048    & 36   & G028.801$+$00.170      &  60   & 102--108     & 107.6      &  0.5    \\
BGPS4933	    & 13   & G032.152$+$00.131      &	60	 & 91--96       & 95         &  0.8    \\
BGPS5539	    & 15   & G035.051$-$00.520    &	52	 & 48--53       & 48         &  0.6    \\
BGPS5821    & 44   & G037.200$-$00.430    &  60   & 34--38       & 38         &  0.6    \\
BGPS5874	    & 9    & G037.820$+$00.414      &	52	 & 12--22       & 22.3       &  0.7    \\
BGPS6418	    & 11   & G053.188$+$00.209      &	53	 & $-$1--2      & 5.3        &  \nodata    \\
BGPS6547	    & 11   & G076.155$-$00.286    &	60	 & $-$32--$-$30 & $-$28.2    &  \nodata    \\    
\enddata
\tablecomments{
Column 1: the maser name. Column 2: the angular separation of KVN targeted center and WISE H~{\sc ii} regions. Columns 3--4: the name and the radius of WISE H~{\sc ii} regions. Column 5: the velocity range of class I methanol maser. Column 6: the RRLs velocity of the H~{\sc ii} regions. Column 7: the peak flux density ratio between 95 and 44 GHz methanol masers.
}
\end{deluxetable}


\begin{thebibliography}{}

\bibitem[Anderson et al.(2014)]{2014ApJS..212....1A} Anderson, L.~D., Bania, T.~M., Balser, D.~S., et al.\ 2014, \apjs, 212, 1 
\bibitem[Anderson et al.(2015)]{2015ApJ...810...42A} Anderson, L.~D., Hough, L.~A., Wenger, T.~V., Bania, T.~M., \& Balser, D.~S.\ 2015, \apj, 810, 42 
\bibitem[Anderson et al.(2018)]{2018ApJS..234...33A} Anderson, L.~D., Armentrout, W.~P., Luisi, M., et al.\ 2018, \apjs, 234, 33 
\bibitem[Bae et al.(2011)]{2011ApJS..196...21B} Bae, J.-H., Kim, K.-T., Youn, S.-Y., et al.\ 2011, \apjs, 196, 21 
\bibitem[Bachiller et al.(1990)]{1990AA...240..116B} Bachiller, R., Gomez-Gonzalez, J., Barcia, A., \& Menten, K.~M.\ 1990, \aap, 240, 116 
\bibitem[Batrla et al.(1987)]{1987Natur.326...49B} Batrla, W., Matthews, H.~E., Menten, K.~M., \& Walmsley, C.~M.\ 1987, \nat, 326, 49 
\bibitem[Breen et al.(2019)]{2019MNRAS.484.5072B} Breen, S.~L., Contreras, Y., Dawson, J.~R., et al.\ 2019, \mnras, 484, 5072 
\bibitem[Breen et al.(2010)]{2010MNRAS.401.2219B} Breen, S.~L., Ellingsen, S.~P., Caswell, J.~L., \& Lewis, B.~E.\ 2010, \mnras, 401, 2219 
\bibitem[Caswell et al.(2010)]{2010MNRAS.404.1029C} Caswell, J.~L., Fuller, G.~A., Green, J.~A., et al.\ 2010, \mnras, 404, 1029 
\bibitem[Chen et al.(2011)]{2011ApJS..196....9C} Chen, X., Ellingsen, S.~P., Shen, Z.-Q., Titmarsh, A., \& Gan, C.-G.\ 2011, \apjs, 196, 9
\bibitem[Chen et al.(2012)]{2012ApJS..200....5C} Chen, X., Ellingsen, S.~P., He, J.-H., et al.\ 2012, \apjs, 200, 5 
\bibitem[Chen et al.(2013)]{2013ApJS..206....9C} Chen, X., Gan, C.-G., Ellingsen, S.~P., et al.\ 2013, \apjs, 206, 9
\bibitem[Choi et al.(2012)]{2012ApJ...759..136C} Choi, M., Kang, M., Byun, D.-Y., \& Lee, J.-E.\ 2012, \apj, 759, 136 
\bibitem[Choi et al.(2014)]{2014ApJ...790...99C} Choi, Y.~K., Hachisuka, K., Reid, M.~J., et al.\ 2014, \apj, 790, 99 
\bibitem[Connelley et al.(2007)]{2007AJ....133.1528C} Connelley, M.~S., Reipurth, B., \& Tokunaga, A.~T.\ 2007, \aj, 133, 1528 
\bibitem[Cragg et al.(1992)]{1992MNRAS.259..203C} Cragg, D.~M., Johns, K.~P., Godfrey, P.~D., \& Brown, R.~D.\ 1992, \mnras, 259, 203
\bibitem[Cragg et al.(2005)]{2005MNRAS.360..533C} Cragg, D.~M., Sobolev, A.~M., \& Godfrey, P.~D.\ 2005, \mnras, 360, 533
\bibitem[Cyganowski et al.(2007)]{2007AJ....134..346C} Cyganowski, C.~J., Brogan, C.~L., \& Hunter, T.~R.\ 2007, \aj, 134, 346
\bibitem[Cyganowski et al.(2008)]{2008AJ....136.2391C} Cyganowski, C.~J., Whitney, B.~A., Holden, E., et al.\ 2008, \aj, 136, 2391-2412 
\bibitem[Deharveng et al.(2008)]{2008A&A...482..585D} Deharveng, L., Lefloch, B., Kurtz, S., et al.\ 2008, \aap, 482, 585
\bibitem[Ellingsen(2005)]{2005MNRAS.359.1498E} Ellingsen, S.~P.\ 2005, \mnras, 359, 1498
\bibitem[Ellingsen(2006)]{2006ApJ...638..241E} Ellingsen, S.~P.\ 2006, \apj, 638, 241
\bibitem[Fontani et al.(2010)]{2010AA...517A..56F} Fontani, F., Cesaroni, R., \& Furuya, R.~S.\ 2010, \aap, 517, A56
\bibitem[Gan et al.(2013)]{2013ApJ...763....2G} Gan, C.-G., Chen, X., Shen, Z.-Q., Xu, Y., \& Ju, B.-G.\ 2013, \apj, 763, 2
\bibitem[Gildas Team(2013)]{2013ascl.soft05010G} Gildas Team\ 2013, GILDAS: Grenoble Image and Line Data Analysis Software, ascl:1305.010
\bibitem[G{\'o}mez-Ruiz et al.(2016)]{2016ApJS..222...18G} G{\'o}mez-Ruiz, A.~I., Kurtz, S.~E., Araya, E.~D., Hofner, P., \& Loinard, L.\ 2016, \apjs, 222, 18
\bibitem[Green et al.(2010)]{2010MNRAS.409..913G} Green, J.~A., Caswell, J.~L., Fuller, G.~A., et al.\ 2010, \mnras, 409, 913 
\bibitem[Haschick \& Baan(1989)]{1989ApJ...339..949H} Haschick, A.~D., \& Baan, W.~A.\ 1989, \apj, 339, 949 
\bibitem[Haschick et al.(1990)]{1990ApJ...354..556H} Haschick, A.~D., Menten, K.~M., \& Baan, W.~A.\ 1990, \apj, 354, 556 
\bibitem[Hunter(2007)]{2007CSE.....9...90H} Hunter, J.~D.\ 2007, Computing in Science and Engineering, 9, 90
\bibitem[Jordan et al.(2015)]{2015MNRAS.448.2344J} Jordan, C.~H., Walsh, A.~J., Lowe, V., et al.\ 2015, \mnras, 448, 2344
\bibitem[Jordan et al.(2017)]{2017MNRAS.471.3915J} Jordan, C.~H., Walsh, A.~J., Breen, S.~L., et al.\ 2017, \mnras, 471, 3915 
\bibitem[Kalenskii et al.(1992)]{1992AZh....69.1002K} Kalenskij, S.~V., Bachiller, R., Berulis, I.~I., et al.\ 1992, \azh, 69, 1002 
\bibitem[Kalenskii et al.(2013)]{2013ARep...57..120K} Kalenskii, S.~V., Kurtz, S., \& Bergman, P.\ 2013, Astronomy Reports, 57, 120 
\bibitem[Kalenskii et al.(1994)]{1994AAS..103..129K} Kalenskii, S.~V., Liljestroem, T., Val'tts, I.~E., et al.\ 1994, \aaps, 103, 129
\bibitem[Kalenskii et al.(2001)]{2001ARep...45...26K} Kalenskii, S.~V., Slysh, V.~I., Val'tts, I.~E., Winnberg, A., \& Johansson, L.~E.\ 2001, Astronomy Reports, 45, 26
\bibitem[Kalenskii et al.(2006)]{2006ARep...50..289K} Kalenskii, S.~V., Promyslov, V.~G., Slysh, V.~I., Bergman, P., \& Winnberg, A.\ 2006, Astronomy Reports, 50, 289 
\bibitem[Kang et al.(2015)]{2015ApJS..221....6K} Kang, H., Kim, K.-T., Byun, D.-Y., Lee, S., \& Park, Y.-S.\ 2015, \apjs, 221, 6
\bibitem[Kang et al.(2016)]{2016ApJS..227...17K} Kang, J.-. hyun ., Byun, D.-Y., Kim, K.-T., et al.\ 2016, \apjs, 227, 17
\bibitem[Kim et al.(2018)]{2018ApJS..236...31K} Kim, C.-H., Kim, K.-T., \& Park, Y.-S.\ 2018, \apjs, 236, 31 
\bibitem[Kim et al.(2019)]{2019ApJS..244....2K} Kim, W.-J., Kim, K.-T., \& Kim, K.-T.\ 2019, \apjs, 244, 2 
\bibitem[Kurtz et al.(1994)]{1994ApJS...91..659K} Kurtz, S., Churchwell, E., \& Wood, D.~O.~S.\ 1994, \apjs, 91, 659 
\bibitem[Kurtz et al.(2004)]{2004ApJS..155..149K} Kurtz, S., Hofner, P., \& {\'A}lvarez, C.~V.\ 2004, \apjs, 155, 149 
\bibitem[Leurini et al.(2016)]{2016A&A...592A..31L} Leurini, S., Menten, K.~M., \& Walmsley, C.~M.\ 2016, \aap, 592, A31
\bibitem[Liechti \& Wilson(1996)]{1996AA...314..615L} Liechti, S., \& Wilson, T.~L.\ 1996, \aap, 314, 615 
\bibitem[Lim et al.(2012)]{2012AJ....144..151L} Lim, W., Lyo, A.-R., Kim, K.-T., \& Byun, D.-Y.\ 2012, \aj, 144, 151
\bibitem[Ladeyschikov et al.(2019)]{2019AJ....158..233L} Ladeyschikov, D.~A., Bayandina, O.~S., \& Sobolev, A.~M.\ 2019, \aj, 158, 233
\bibitem[McCarthy et al.(2018)]{2018MNRAS.477..507M} McCarthy, T.~P., Ellingsen, S.~P., Voronkov, M.~A., \& Cim{\`o}, G.\ 2018, \mnras, 477, 507
\bibitem[Menten(1991a)]{1991ASPC...16..119M} Menten, K.\ 1991a, Atoms, Ions and Molecules: New Results in Spectral Line Astrophysics, 16, 119
\bibitem[Menten(1991b)]{1991ApJ...380L..75M} Menten, K.~M.\ 1991b, \apjl, 380, L75 
\bibitem[Miralles et al.(1997)]{1997ApJ...488..749M} Miralles, M.~P., Salas, L., Cruz-Gonz{\'a}lez, I., \& Kurtz, S.\ 1997, \apj, 488, 749 
\bibitem[Molinari et al.(1996)]{1996AA...308..573M} Molinari, S., Brand, J., Cesaroni, R., et al.\ 1996, \aap, 308, 573
\bibitem[Molinari et al.(1998)]{1998AA...336..339M} Molinari, S., Brand, J., Cesaroni, R., Palla, F., \& Palumbo, G.~G.~C.\ 1998, \aap, 336, 339 
\bibitem[Morimoto et al.(1985)]{1985ApJ...288L..11M} Morimoto, M., Kanzawa, T., \& Ohishi, M.\ 1985, \apjl, 288, L11
\bibitem[Motte et al.(2018)]{2018ARAA..56...41M} Motte, F., Bontemps, S., \& Louvet, F.\ 2018, \araa, 56, 41 
\bibitem[Pandian et al.(2007)]{2007ApJ...656..255P} Pandian, J.~D., Goldsmith, P.~F., \& Deshpande, A.~A.\ 2007, \apj, 656, 255 
\bibitem[Peretto \& Fuller(2009)]{2009AA...505..405P} Peretto, N., \& Fuller, G.~A.\ 2009, \aap, 505, 405 
\bibitem[Peretto et al.(2016)]{2016AA...590A..72P} Peretto, N., Lenfestey, C., Fuller, G.~A., et al.\ 2016, \aap, 590, A72 
\bibitem[Pestalozzi et al.(2005)]{2005AA...432..737P} Pestalozzi, M.~R., Minier, V., \& Booth, R.~S.\ 2005, \aap, 432, 737 
\bibitem[Pety(2005)]{2005sf2a.conf..721P} Pety, J.\ 2005, SF2A-2005: Semaine De L'astrophysique Francaise, 721
\bibitem[Pomar{\`e}s et al.(2009)]{2009A&A...494..987P} Pomar{\`e}s, M., Zavagno, A., Deharveng, L., et al.\ 2009, \aap, 494, 987
\bibitem[Pillai et al.(2006)]{2006AA...450..569P} Pillai, T., Wyrowski, F., Carey, S.~J., \& Menten, K.~M.\ 2006, \aap, 450, 569  
\bibitem[Pratap et al.(2008)]{2008AJ....135.1718P} Pratap, P., Shute, P.~A., Keane, T.~C., Battersby, C., \& Sterling, S.\ 2008, \aj, 135, 1718
\bibitem[Rathborne et al.(2006)]{2006ApJ...641..389R} Rathborne, J.~M., Jackson, J.~M., \& Simon, R.\ 2006, \apj, 641, 389 
\bibitem[Reid et al.(2014)]{2014ApJ...783..130R} Reid, M.~J., Menten, K.~M., Brunthaler, A., et al.\ 2014, \apj, 783, 130
\bibitem[Robitaille, \& Bressert(2012)]{2012ascl.soft08017R} Robitaille, T., \& Bressert, E.\ 2012, APLpy: Astronomical Plotting Library in Python, ascl:1208.017
\bibitem[Rodr{\'{\i}}guez-Garza et al.(2017)]{2017ApJS..233....4R} Rodr{\'{\i}}guez-Garza, C.~B., Kurtz, S.~E., G{\'o}mez-Ruiz, A.~I., et al.\ 2017, \apjs, 233, 4 
\bibitem[Rosolowsky et al.(2010)]{2010ApJS..188..123R} Rosolowsky, E., Dunham, M.~K., Ginsburg, A., et al.\ 2010, \apjs, 188, 123-138 
\bibitem[Rygl et al.(2010)]{2010AA...511A...2R} Rygl, K.~L.~J., Brunthaler, A., Reid, M.~J., et al.\ 2010, \aap, 511, A2
\bibitem[Shepherd, \& Churchwell(1996)]{1996ApJ...472..225S} Shepherd, D.~S., \& Churchwell, E.\ 1996, \apj, 472, 225
\bibitem[Simon et al.(2006)]{2006ApJ...639..227S} Simon, R., Jackson, J.~M., Rathborne, J.~M., \& Chambers, E.~T.\ 2006, \apj, 639, 227
\bibitem[Slysh et al.(1994)]{1994MNRAS.268..464S} Slysh, V.~I., Kalenskii, S.~V., Valtts, I.~E., \& Otrupcek, R.\ 1994, \mnras, 268, 464 
\bibitem[Sobolev \& Deguchi(1994)]{1994A&A...291..569S} Sobolev, A.~M., \& Deguchi, S.\ 1994, \aap, 291, 569
\bibitem[Sobolev \& Parfenov(2018)]{2018IAUS..336...57S} Sobolev, A.~M., \& Parfenov, S.~Y.\ 2018, Astrophysical Masers: Unlocking the Mysteries of the Universe, 336, 57
\bibitem[Sridharan et al.(2002)]{2002ApJ...566..931S} Sridharan, T.~K., Beuther, H., Schilke, P., Menten, K.~M., \& Wyrowski, F.\ 2002, \apj, 566, 931
\bibitem[Ulich \& Haas(1976)]{1976ApJS...30..247U} Ulich, B.~L., \& Haas, R.~W.\ 1976, \apjs, 30, 247
\bibitem[Val'tts et al.(1995)]{1995AZh....72...22V} Val'tts, I.~E., Dzyura, A.~M., Kalenskii, S.~V., et al.\ 1995, \azh, 72, 22 
\bibitem[Val'tts et al.(2000)]{2000MNRAS.317..315V} Val'tts, I.~E., Ellingsen, S.~P., Slysh, V.~I., et al.\ 2000, \mnras, 317, 315
\bibitem[van der Walt et al.(2011)]{2011CSE....13b..22V} van der Walt, S., Colbert, S.~C., \& Varoquaux, G.\ 2011, Computing in Science and Engineering, 13, 22
\bibitem[Voronkov et al.(2014)]{2014MNRAS.439.2584V} Voronkov, M.~A., Caswell, J.~L., Ellingsen, S.~P., Green, J.~A., \& Breen, S.~L.\ 2014, \mnras, 439, 2584
\bibitem[Walsh et al.(1998)]{1998MNRAS.301..640W} Walsh, A.~J., Burton, M.~G., Hyland, A.~R., \& Robinson, G.\ 1998, \mnras, 301, 640
\bibitem[Wang et al.(2011)]{2011AA...527A..32W} Wang, Y., Beuther, H., Bik, A., et al.\ 2011, \aap, 527, A32
\bibitem[Wood \& Churchwell(1989)]{1989ApJ...340..265W} Wood, D.~O.~S., \& Churchwell, E.\ 1989, \apj, 340, 265 
\bibitem[Wright et al.(2010)]{2010AJ....140.1868W} Wright, E.~L., Eisenhardt, P.~R.~M., Mainzer, A.~K., et al.\ 2010, \aj, 140, 1868-1881
\bibitem[Xu et al.(2003)]{2003ChJAA...3...49X} Xu, Y., Zheng, X.-W., \& Jiang, D.-R.\ 2003, \cjaa, 3, 49 
\bibitem[Yan et al.(2018)]{2018MNRAS.476.3981Y} Yan, Q.-Z., Xu, Y., Walsh, A.~J., et al.\ 2018, \mnras, 476, 3981 
\bibitem[Yang et al.(2017)]{2017ApJS..231...20Y} Yang, W., Xu, Y., Chen, X., et al.\ 2017, \apjs, 231, 20
\bibitem[Zhang et al.(2001)]{2001ApJ...552L.167Z} Zhang, Q., Hunter, T.~R., Brand, J., et al.\ 2001, \apjl, 552, L167

\end{thebibliography}
\end{document}